\documentclass[prd, superscriptaddress, tightenlines, longbibliography, nofootinbib, eqsecnum, amsfonts, amsmath, floatfix, notitlepage, twocolumn]{revtex4-2}
\usepackage[utf8]{inputenc}
\usepackage{mathrsfs}
\usepackage{mathtools}
\usepackage{euscript}
\usepackage{graphicx}
\usepackage{amsmath}
\usepackage{amssymb}
\usepackage{gensymb}
\usepackage{bm}
\usepackage[usenames,dvipsnames,svgnames,table]{xcolor}
\definecolor{linkcolor}{rgb}{0.6,0,0}
\definecolor{citecolor}{rgb}{0,0,0.75}
\definecolor{urlcolor}{rgb}{0.12,0.46,0.7}
\usepackage{xspace}
\usepackage{wasysym}
\usepackage{times}
\usepackage{appendix}
\usepackage{comment}
\usepackage{lipsum}
\usepackage[nolist,nohyperlinks]{acronym}
\usepackage{float}
\usepackage{natbib, ifthen}
\usepackage[breaklinks, colorlinks, urlcolor=urlcolor, linkcolor=linkcolor,citecolor=citecolor,pdfencoding=auto]{hyperref}
\usepackage{longtable}
\usepackage{listings}
\usepackage{bbm}
\usepackage{booktabs}
\usepackage{hyperref}
\usepackage{placeins}
\pdfoutput=1

\newcommand{\la}{\langle}

\providecommand{\planck}{\textit{Planck}}
\providecommand{\Planck}{\planck}

\newcommand{\mksym}[1]{\ifmmode {\rm #1}\else #1\fi}

\newcommand{\rdrag}{r_{\rm drag}}
\newcommand{\zstar}{z_{\ast}}
\newcommand{\rstar}{r_{\ast}}

\providecommand{\Planck}{\textit{Planck}}
\providecommand{\planck}{\Planck}

\providecommand{\lea}{\la}
\providecommand{\gea}{\ga}

\providecommand{\alt}{\lea}
\providecommand{\agt}{\gea}
\providecommand{\text}[1]{\rm{#1}}

\providecommand{\CAMB}{{\tt camb}}
\providecommand{\GetDist}{{\tt GetDist}}
\providecommand{\Cobaya}{{\tt Cobaya}}

\providecommand{\LCDM}{{$\rm{\Lambda CDM}$}}

\newcommand{\begm}{\begin{pmatrix}}
\newcommand{\enm}{\end{pmatrix}}

\newcommand\ba{\begin{eqnarray}}
\newcommand\ea{\end{eqnarray}}
\newcommand\bea{\begin{eqnarray}}
\newcommand\eea{\end{eqnarray}}

\newcommand\be{\begin{equation}}
\newcommand\ee{\end{equation}}

\renewcommand{\DH}{D_H}
\newcommand{\DHL}{D_H^\Lambda}
\newcommand{\DM}{D_M}
\newcommand{\DML}{D_M^\Lambda}
\newcommand{\FAP}{F_{\rm AP}}
\newcommand{\FAPL}{F_{\rm AP}^{\Lambda}}

\newcommand{\DV}{D_{V}}
\newcommand{\DVL}{D_V^{\Lambda}}

\newcommand{\isdraft}[1]{}

\begin{document}
\title{Understanding acoustic scale observations: the one-sided fight against $\Lambda$}

\newcommand{\Sussex}{Department of Physics \& Astronomy, University of Sussex, Brighton BN1 9QH, UK}

\author{Antony Lewis}
\homepage{https://cosmologist.info}
\affiliation{\Sussex}

\author{Ewan Chamberlain}
\affiliation{\Sussex}

\date{\today}

\begin{abstract}
The cosmic microwave background (CMB) and baryon acoustic oscillations (BAO) provide precise measurements of the cosmic expansion history through the comoving acoustic scale. The CMB angular scale measurement $\theta_*$ is particularly robust, constraining the ratio of the sound horizon to the angular diameter distance to last scattering independently of the late-time cosmological model. For models with standard early-universe physics, this measurement strongly constrains possible deviations from $\Lambda$CDM at late times. We show that the null energy condition imposes strict inequalities on the BAO observables $D_H(z)$, $D_M(z)$, $D_V(z)$ and $\FAP(z)$ relative to $\Lambda$CDM predictions. These inequalities demonstrate that certain deviations from \LCDM\ are impossible for any physical non-interacting dark energy model that respects the null energy condition within the context of FRW cosmological models. We also identify the regions of parameter space in the CPL parameterization $w(a) = w_0 + w_a(1-a)$ that can give predictions consistent with both the null energy condition and the observed CMB scale. While current DESI DR2 BAO measurements exhibit some joint-constraint parameter tensions with \LCDM, this tension arises primarily in directions that are inconsistent with the null-energy condition, so \LCDM\ is favoured by current acoustic scale measurements unless the null-energy condition is violated.
\end{abstract}

\maketitle

\section{Introduction}

The cosmic microwave background (CMB) and baryon acoustic oscillations (BAO) provide complementary probes of the cosmic expansion history through precise measurements of the comoving acoustic scale. The CMB primarily constrains the angular size of the sound horizon at last scattering through the parameter $\theta_* = r_*/\DM(z_*)$, where $r_*$ is the comoving sound horizon at recombination and $\DM(z_*)$ is the comoving angular diameter distance to the last scattering surface at redshift $z_*$. This measurement is remarkably robust --- the angular scale of the acoustic peaks in the CMB power spectrum is measured to better than 0.03\% precision by \Planck~\cite{PCP2018}, independently of the assumed late-time cosmological model~\cite{Lemos:2023xhs}.

BAO measurements at lower redshift detect the imprint of these same acoustic oscillations in the clustering of galaxies and other tracers of large-scale structure. Modern BAO analyses can measure both the line-of-sight and transverse clustering, allowing separate constraints on the Hubble parameter $H(z)$ and comoving angular diameter distance $\DM(z)$ at the survey redshift. The line-of-sight BAO scale is measured as a redshift interval, which translates to a constraint on  $\rdrag/\DH(z)$ where $\rdrag$ is the comoving sound horizon distance at the drag epoch when baryons decouple from the photons, and $D_H(z) = c/H(z)$ [from now on we use natural units where $c=1$]. The transverse BAO scale measures the angular scale $\rdrag/\DM(z)$. 

When analysing anisotropic clustering, it is common to also report constraints on the combination $D_V(z) = [z\DM^2(z)\DH(z)]^{1/3}$ which determines the angle-averaged BAO scale, and the Alcock-Paczynski parameter $\FAP(z) = \DM(z)/\DH(z)$ which quantifies the ratio of the transverse and radial clustering scales (and is independent of $\rdrag$). The latest Dark Energy Spectroscopic Instrument (DESI) BAO analysis constrains these parameters (or $\DM$, $\DH$) to approximately 1\% precision over a range of redshifts in the range $0.2 \alt z \alt 2$~\cite{DESI:2024mwx,DESI:2025zgx}.

For fixed early-universe physics (and hence fixed $\rdrag$ and $r_*$), the CMB acoustic scale measurement essentially fixes $D_M(z_*)$. Combined with BAO measurements of $\DM(z)$ and $\DH(z)$ at lower redshift, this provides powerful constraints on the expansion history. Recent analyses of DESI and other BAO data have suggested possible tensions with \LCDM, with some preference for thawing-like behaviour at low redshift and 
`phantom' dark energy with equation of state parameter $w < -1$ at high redshift~\cite{DESI:2024aqx,DESI:2024kob,Cortes:2024lgw,Linder:2024rdj,Fikri:2024klc,DESI:2025zgx,DESI:2025fii}. However, models that violate the null energy condition (NEC) are likely to be unphysical, or at least require more radical modification to our current understanding of physics. It is therefore important to understand whether the data are consistent with the null-energy condition or not~\cite{Sen:2007ep}. 

The null energy condition requires that the energy density of any non-interacting physical fluid cannot increase as the universe expands. The energy density must therefore be constant or increase with redshift. For dark energy, this implies $d\rho_{\rm de}/dz \geq 0$. However, any attempt to deviate from a cosmological constant $\Lambda$ (with a constant dark energy density) at low redshift must have a compensating change at higher redshift to ensure that the observed scale of the CMB acoustic peaks determined by $\DM(z_*)$ remains consistent.  This balancing act strongly constrains the allowed behaviour of $H(z)$, $\DM(z)$ and derived quantities like $D_V(z)$ and $\FAP(z)$. These constraints become particularly relevant when interpreting apparent deviations from \LCDM\ with current and forthcoming BAO data. In the $\Lambda$CDM model, where the dark energy is a cosmological constant, the null-energy bound is saturated (the density is constant). The null-energy condition therefore imposes a one-sided constraint, which, as we shall show, strongly restricts the directions in which $\DH(z)$ and $\DM(z)$ can deviate from \LCDM\ while remaining consistent with the null-energy condition.

A common parameterization used to explore dynamical dark energy models is the CPL form~\cite{Chevallier:2000qy,Linder:2002et}:
\begin{equation}
w(a) = w_0 + w_a(1-a),
\end{equation}
where $w_0$ is the equation of state parameter today and $w_a$ describes its evolution with scale factor $a$. 
This gives a simple non-physical parameterization that can be easily compared with data, but interpreting the results can be difficult in terms of physical models. It is possible for a model to satisfy the
null-energy condition, but, when constrained by measurements over a finite low-redshift range, to give constraints in the  $(w_0,w_a)$ plane that naively imply a violation of the null-energy condition (phantom behaviour) at high redshift~\cite{Shlivko:2024llw,Linder:2024rdj}. Although the high-redshift behaviour is not directly constrained by low-redshift observations, there is a strong integral constraint that the angular scale measured in the CMB remains at the observed angular size. Imposing this constraint, the deviation of the BAO observables from \LCDM\ can be qualitatively different (e.g. in sign) if the null-energy condition is imposed compared to the predictions for the naive $w_0, w_a$ model. We identify the regions of $w_0, w_a$ parameter space that can be consistent with the null-energy bounds, generalizing previous results for specific (e.g. thawing~\cite{Shlivko:2024llw,Wolf:2024eph,Payeur:2024dnq}) models. 
 Understanding which regions are physically allowed  is crucial for interpreting any deviations from \LCDM\ indicated by current and forthcoming data.

In Sec.~\ref{sec:constraints} we first derive the constraints on observable quantities imposed by the null energy condition when the observed CMB acoustic scale and other non-dark energy parameters are held fixed. We show that these constraints rule out substantial regions of possible observation space. Then in Sec.~\ref{sec:CPL} we show how these translate into an interpretation of allowed regions of the $w_0$, $w_a$ (CPL) parameter space. In Sec.~\ref{sec:DESI} we review
current BAO measurements in light of these results, showing that the BAO data and CMB constraints favour \LCDM\ over other null-energy consistent models.

\section{Constraints from the null-energy condition}
\label{sec:constraints}

We assume a flat FRW universe, with a non-interacting dark energy component (which either is or is not a cosmological constant), plus other matter and radiation. CMB measurements tightly constrain the CMB temperature $T_{\rm CMB}$, angular scale $\theta_*$, and physical density parameters $\Omega_m h^2$, $\Omega_b h^2$. Given the tight constraints, we could approximate these parameters as fixed, with Helium abundances determined by standard big-bang nucleosynthesis.
More generally, we can consider any point in parameter space with a given set of cosmological parameters, and compare the late-time evolution obtained using different dark energy models. We choose to focus on parameterizations where $\theta_*$ is used, so that $H_0$ or the dark energy density today are derived parameters that can vary when varying the dark energy model at fixed $\theta_*$. 
We then ask the question: For a given set of non-dark energy parameters, how do the late-time evolution and observables vary depending on the dark-energy model?

We assume that dark energy is negligible at high redshift, so it has no effect on the CMB or baryons at the time of recombination or baryon decoupling. 
For a given set of non-dark-energy parameters, the comoving acoustic scale as parameterized by $\rdrag$ or $\rstar$ is then fixed. For given fixed $\theta_* = \rstar/\DM(z_*)$, the comoving distance to last scattering, $\DM(z_*)$ is then also fixed.
The comoving distance is given by
    \begin{equation}
    \DM(z) = \int_{0}^{z} \frac{dz'}{H(z')}= \int_{0}^{z} \DH(z') \, dz',
    \end{equation}
and hence $\DH$ and $\DM$ are also related by
    \begin{equation}
    \DH = \frac{d \DM}{dz}, \quad \DHL = \frac{d\DML}{dz}.
    \end{equation}
    
For fixed $\theta_*$, we therefore have the integral constraint ensuring that the angular diameter distance must be the same in the two dark energy models:
    \begin{equation}
    \DM(\zstar) = \int_{0}^{\zstar} \DH \, dz = \int_{0}^{\zstar} \DHL \, dz = \DML(\zstar).
    \label{eq:intconstraint}
    \end{equation}
 
The evolution of the Hubble parameter is determined by the Friedmann equation    
    \begin{equation}
    D_H^{-2} = H^2 = \frac{8\pi G}{3}\left(\rho_m  + \rho_{\rm de}\right),
    \end{equation}
where here we slightly abuse notation so that $\rho_m$ is taken to include all non-dark energy components (matter, radiation, massive neutrinos, and, if needed, curvature\footnote{We only show results explicitly for a flat universe, but the results in Table~\ref{table:ineq} also hold as long as the curvature constant $K\propto \Omega_K h^2$ is held fixed as part of the non dark-energy parameters (assuming a non-extreme model, so the comoving angular diameter distance remains a monotonic function of the comoving radial distance).}).
For fixed non-dark energy parameters $\rho_m(z)$ is a known fixed function.

The null energy condition applied to a dark energy fluid with pressure $p_{\rm de}$ states that $p_{\rm de}\ge -\rho_{\rm de}$, a constraint that is saturated for a cosmological constant.
In an expanding universe, conservation of stress-energy for any NEC-consistent non-interacting dark energy fluid then implies that the density cannot increase with the expansion, hence, in terms of redshift,
    \begin{equation}    
    \frac{d\rho_{\rm de}^\Lambda}{dz} =0,
    \qquad
    \frac{d\rho_{\rm de}}{dz} \ge 0.    
    \end{equation}
 Since $\rho_m(z)$ is fixed, we can combine this with the Friedmann equation to get a derivative constraint on $\DH$:
    \begin{equation}
    \frac{d \DH}{dz} \le \left( \frac{\DH}{\DHL} \right)^3 \frac{d \DHL}{dz}.
    \label{eq:hubblederiv}
    \end{equation}
We assumed that the dark energy density today is positive, as implied by observations. For a homogeneous non-interacting background dark fluid, consistency with the null-energy condition is then equivalent to consistency with the weak energy condition (which for the dark energy fluid additionally stipulates that $\rho_{\rm de}\ge 0$).

\begin{table*}[htbp]
\centering
\begin{tabular}{lll}
\toprule
\textbf{Inequality} & \textbf{Redshift} & \textbf{Notes} \\
\midrule
$\displaystyle \frac{d\DH}{dz} \leq \left( \frac{\DH}{\DHL} \right)^3 \frac{d\DHL}{dz}$  & For all $z$ & Friedmann+null energy condition. \\[2ex]
$\displaystyle \DH(0) \geq \DHL(0)$ & At $z = 0$ & $H_0 < H_0^\Lambda$ today for fixed $\theta_*$. \\[2ex]
$\displaystyle \DH(z) \geq \DHL(z)$ & For $0 \leq z \leq z_c$ & $\DH$ and $\DHL$ cross at $z = z_c$. \\[2ex]
$\displaystyle \DH(z) \leq \DHL(z)$ & For $z_c \leq z \leq \zstar$ & Only cross once, for high $z$: $\DH(z) \rightarrow \DHL(z)$. \\[2ex]
$\displaystyle \DM(z) \geq \DML(z)$ & For all $z$ & Comoving distance inequality holds universally. \\[2ex]
$\displaystyle \frac{d}{dz}\left( \frac{\DM}{\DML} \right) \le 0$ & For all $z$ & Monotonic decrease of $\DM / \DML$ towards 1 as $z$ increases. \\[2ex]
$\displaystyle \frac{\DH}{\DHL} \leq \frac{\DM}{\DML}$ & For all $z$ & Tends to equality at $z \in (0, z_*)$. \\[2ex]
$\displaystyle \FAP \geq \FAPL$ & For all $z$ & Tends to equality at $z \in (0, z_*)$. \\[2ex]
$\displaystyle \DV \geq \DVL$ & For $0\le z \le z_c$ & At $z=0$, $\DV(0)/\DVL(0)=\DH(0)/\DHL(0) > 1$. \\[2ex]
$\displaystyle \frac{\DV}{\DVL} \geq \left( \frac{\FAP}{\FAPL} \right)^{-1/3}$ & For $z > z_c$ & Goes slightly below one at some point at $z>z_c$ \\
\bottomrule
\end{tabular}
\caption{Inequalities for given fixed non-dark energy parameters and $\theta_*$, assuming the null-energy condition and negligible dark energy at recombination ($z=z_*$).}
\label{table:ineq}
\end{table*}

\begin{figure*}[htbp]
    \centering
    \includegraphics[width=0.9\textwidth]{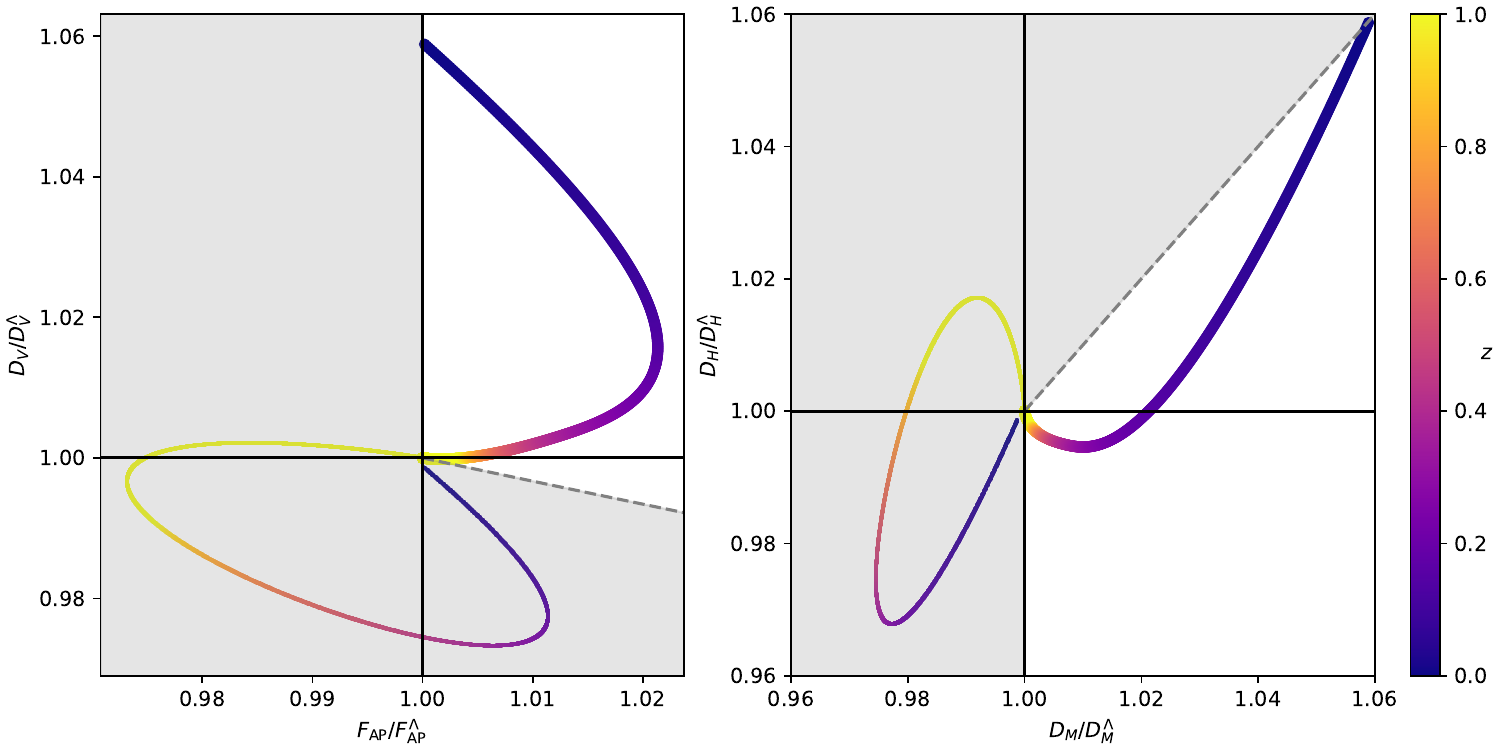}
    \caption{Null-energy condition exclusion regions from Table~\ref{table:ineq} are shown in gray, compared to the predicted evolution for a typical thawing model (thick line) and a $w_0$, $w_a$ model with $w_a=-4(1+w_0)$ and $w_0=-0.7$ (thin line). Lines are coloured according to the corresponding redshift of each point, and colour scale saturates at $z=1$. 
    By construction, the thawing model is consistent with the null-energy condition at all redshifts, but this $w_0, w_a$ model is consistent at none.
    Note that being outside the exclusion region is necessary but not sufficient for a full path to be consistent with the null-energy condition, given the other redshift-dependent constraints in Table~\ref{table:ineq}.  The reference $\Lambda$CDM model is the Planck 2018 best fit.
    }
    \label{fig:BAO-null-regions}
\end{figure*}

\clearpage

We can now derive a set of inequalities that the null energy condition imposes on the parameters measured by BAO measurements at low redshift, either parameterized by the pair $\DM$, $\DH$ or by the pair $\DV$, $\FAP$. The results are summarized in Table~\ref{table:ineq} for convenience.

\subsection{Properties of $D_H$}
The null energy condition implies energy density can only increase with redshift, so the Friedmann equation then implies $\DH(z)$ and $\DHL(z)$ are both monotonically decreasing for $z > 0$.
    
\subsubsection{Hubble parameter today}

Since the comoving distance to the last scattering surface $\DM(z_*)$ is fixed (due to the fixed $\theta_*$) in both models, Eq.~\eqref{eq:intconstraint} combined with the Friedmann equation implies:
\begin{equation}
    \int_{0}^{z_*} \frac{dz}{\sqrt{\rho_m(z) + \rho_{\rm de}(z)}}  = \int_{0}^{z_*} \frac{dz}{\sqrt{\rho_m(z) + \rho_\Lambda}}.
\end{equation}
The null energy condition requires that the dark energy density increases with redshift, which implies that $\rho_{\mathrm{de}}(z) \geq \rho_{\mathrm{de}}(0)$ for $z \geq 0$. Hence, to have the same integral for given $\rho_m$, we must have $\rho_{\rm de}(0)<\rho_\Lambda$. The Friedmann equation then implies that $H_0<H_0^\Lambda$ and hence
\begin{equation}
\DH(0) \ge \DHL(0).
\end{equation}
     
Deviations from $\Lambda$ that satisfy the null energy condition therefore potentially exacerbate any \LCDM\ Hubble tension between the high value of $H_0$ from forward distance ladder measurements~\cite{Riess:2022mme,Freedman:2024eph,Riess:2024vfa,Murakami:2023xuy,Breuval:2024lsv} and the lower value from \Planck\ (and other acoustic measurements). However, the forward distance ladder measurements are really based on calibrated luminosity distances to moderate redshift, not redshift zero, so to reach a firm conclusion we first need to understand how the inferred integrated distances change.
    
\subsubsection{Intersection point}

The integral constraint that $\DM(z_*) = \DML(z_*)$  (Eq.~\ref{eq:intconstraint}) implies that  \(\DH(z)\) and \(\DHL(z)\) 
 must cross at some redshift \(z_c\):
\begin{equation}
\DH(z_c) = \DHL(z_c) \quad \text{for some} \quad 0 < z_c < \zstar.
\end{equation}
 Physically, this crossing point corresponds to the redshift where the dark energy densities in both models are equal, \(\rho_{\rm de}(z_c) = \rho_\Lambda\).
The null energy condition then implies that $\DH<\DHL$ at all higher redshifts (this also follows mathematically from Eq.~\eqref{eq:hubblederiv}). Hence, there is only one crossing and we have
\begin{align}
\DH \geq \DHL   & \quad \text{for } z \leq z_c, \\
\DH \leq \DHL   & \quad \text{for } z \geq z_c. \label{DHhigh}
\end{align}
The ratio $\DH/\DHL$ starts above one at $z=0$, and monotonically decreases until it crosses at $z=z_c$. It then goes below one and gradually asymptotes to one again at high redshift where the densities are assumed to be equal.

\subsection{Properties of $\DM$}

\subsubsection{\(\DM(z) \geq \DML(z)\)}
\label{sec:DMDML}
Near zero redshift, $\DM$ approaches zero, but by l'Hôpital's rule for the ratio we have
\begin{equation}
\frac{\DM(0)}{\DML(0)} \rightarrow \frac{\DH(0)}{\DHL(0)} > 1.
\end{equation}
At low redshift before the crossing point, \(z < z_c\), we have \(\DH(z) > \DHL(z)\), which implies that
\begin{equation}
\DM(z) = \int_{0}^{z} \DH(z') \, dz' > \int_{0}^{z} \DHL(z') \, dz' = \DML(z).
\end{equation}
Therefore, at $z<z_c$, we have $\DM/\DML \ge 1$.
For higher redshifts where $z\ge z_c$ we have $\DH < \DHL$, so 
\begin{equation}
  \frac{d}{dz}\left( \DM-\DML \right) = \DH -\DHL < 0.
\end{equation}
Hence $ \DM-\DML$ monotonically decreases from a positive value at $z_c$ to zero at $z_*$.
We conclude that
\begin{equation}
\DM(z) \geq \DML(z) \quad \text{for all} \quad z > 0.
\end{equation}

Since the luminosity distance $D_L(z)$ at a given redshift is just proportional to $\DM(z)$, and the reported forward distance ladder value for $H_0\propto 1/D_L$, this is now sufficient to show that null-energy consistent dark energy can only exacerbate the Hubble tension.

\subsubsection{\(\frac{\DH}{\DHL} \leq \frac{\DM}{\DML}\)}

For high redshifts, where $z\ge z_c$, we know from Eq.~\eqref{DHhigh} that $\DH/\DHL \leq 1$. Hence, since $\DM > \DML$, the inequality in the subtitle is immediately satisfied. For lower redshifts where $z\le z_c$, define the ratio
\begin{equation}
R(z) \equiv \frac{\DH}{\DHL}.
\end{equation}

The comoving distances are given by
\begin{equation}
\DM(z) = \int_{0}^{z} \DH(z') \, dz', \quad \DML(z) = \int_{0}^{z} \DHL(z') \, dz'.
\end{equation}

We can express \(\DM(z)\) in terms of \( R(z) \) and \(\DHL(z)\) as
\begin{equation}
\DM(z) = \int_{0}^{z} R(z') \DHL(z') \, dz'.
\end{equation}

Since \( R(z) \) is a decreasing function for \( z \leq z_c \) (as shown in Appendix~\ref{appendix:proof}), and \(\DHL(z') > 0\), the weighted average of \( R(z') \) over the interval \([0, z]\) satisfies
\begin{equation}
\frac{\DM(z)}{\DML(z)} = \frac{\int_{0}^{z} R(z') \DHL(z') \, dz'}{\int_{0}^{z} \DHL(z') \, dz'} \geq R(z) \qquad [z \le z_c].
\end{equation}
We conclude that
\begin{equation}
\frac{\DH}{\DHL} \leq \frac{\DM}{\DML} \qquad \text{for all } z.
\end{equation}

\subsubsection{Monotonic decrease of \( \DM(z) / \DML(z) \) }
This result also implies that the ratio \( \DM(z) / \DML(z) \) is a decreasing function of \( z \) at all times: 
\begin{equation}
\frac{d}{dz}\left(\frac{\DM}{\DML}\right) = \frac{\DH}{{\DML}} - \frac{\DM}{(\DML)^2}\DHL < 0.
\end{equation}
Hence, \( \DM(z) / \DML(z) \) decreases monotonically towards one at high redshift.

\subsection{Properties of $\FAP \equiv \frac{\DM}{\DH}$} 
\label{sec:FAP}

Since $\frac{\DH}{\DHL} \leq \frac{\DM}{\DML}$, it follows immediately that $\FAP \ge \FAPL$ for all $z$. From the limiting behaviours at high and low redshift, $\FAP/\FAPL$ increases from one at $z=0$ and then decreases back to one at $z=z_*$.

\subsection{Properties of $\DV \equiv \left(z \DM^2 \DH\right)^{1/3}$}
\label{sec:DV}
At low redshift, the limiting behaviour as $z\rightarrow 0$ is  $\DV(z)/\DVL(z) \rightarrow \DH/\DHL \ge 1$. 
For low redshift where $z\le z_c$, the constraints on $\DM$ and $\DH$ immediately imply that $\DV/\DVL \ge 1$.
Writing
\begin{equation}
\frac{\DV}{\DVL} = \frac{ d (\DM{}^3)}{d(\DML{}^3)},
\end{equation}
this cubic derivative starts larger than one at $z=0$. However, we know $\DM(0)=\DML(0)=0$ and that at high redshift $\DM(z_*)^3 = \DML(z_*)^3$, so we infer that the derivative must cross below one at some point if it starts above one.
Hence, $\DV/\DVL$ must cross below one at some point at $z>z_c$. 

While $\DV/\DVL$ is not always larger than one, it typically never gets much lower: since $\DM\ge \DML$ we have
\begin{equation}
\frac{\DV}{\DVL} \ge \left(\frac{\DM}{\DML}\right)^{1/3} \left(\frac{\DH}{\DHL}\right)^{1/3}.
\label{eq:DVlim}
\end{equation}
Since $\DH/\DHL$ tends to one from below in matter domination, $\DV/\DVL$ is always at least three times closer to one at high redshift than $\DH/\DHL$. Hence, generically, $\DV>\DVL$ (and always at low redshift), or $\DV$ is close to $\DVL$.

Quantitatively, we can use
\begin{equation}
\frac{\DH}{\DHL} = \frac{\DV}{\DVL} \left(\frac{\FAP}{\FAPL}\right)^{-2/3}, 
\qquad
\frac{\DM}{\DML} = \frac{\DV}{\DVL}\ \left(\frac{\FAP}{\FAPL}\right)^{1/3}. 
\end{equation}
In terms of $\FAP$ the bound of Eq.~\eqref{eq:DVlim} for high redshift, $z>z_c$, then becomes
\begin{equation}
\frac{\DV}{\DVL} \ge \left(\frac{\FAP}{\FAPL}\right)^{-1/3}.
\end{equation}

\subsection{Uncalibrated supernovae}

Although our focus here is on acoustic scale measurements, similar results can be derived for other observables.
For example, without knowing the absolute calibration standardized supernovae observations constrain ratios of luminosity distances, and hence ratios $\DM(z_2)/\DM(z_1)$. From the monotonicity of $\DM/\DML$ we have the monotonicity requirement for all $z_1$, $z_2$:
\begin{equation}
\frac{\DM(z_2)}{\DM(z_1)}\frac{\DML(z_1)}{\DML(z_2)} \le 1 \qquad \text{for } z_2\ge z_1.
\end{equation}
This monotonicity constraint is consistent with the direction of a possible low-redshift deviation from \LCDM\ discussed by Refs.~\cite{DES:2024jxu,Rubin:2023ovl,DESI:2024kob}. 
As mentioned in Sec.~\ref{sec:DMDML}, this would, however, potentially exacerbate the tension in the distance ladder calibration. As we shall see, such a deviation is also harder to reconcile with high-redshift and acoustic scale observations while remaining consistent with the null-energy condition.

\section{Predictions from CPL}
\label{sec:CPL}

For a given pair of $w_0$, $w_a$ parameters, the linear CPL $w=w_0+(1-a)w_a$ model defines the entire evolution of the dark energy density. The null-energy condition requires that $w\ge -1$ at all times, which will be violated at high redshift if $w_a < -(1+w_0)$. However, in practice no-one seriously thinks CPL is a physical model, so it is instead often interpreted a way to parameterize deviations from \LCDM\ over some finite redshift range. Having  $w_a < -(1+w_0)$ does not necessarily prevent predictions for observables that are consistent with the null energy condition, especially if the observables are at relatively low redshift. Indeed, it is straightforward to construct (null-energy consistent) thawing quintessence models that give observables at $z\alt 1$ that are consistent to good accuracy with the predictions of a CPL model that appear to violate this bound~\cite{dePutter:2008wt,Wolf:2023uno,Linder:2024rdj,Shlivko:2024llw}. However, this does not mean that the entire $w_0$, $w_a$ plane can give predictions for observables that are consistent with the null-energy condition, which we investigate further in this section.

\begin{figure}[]
    \centering
    \includegraphics[width=\columnwidth]{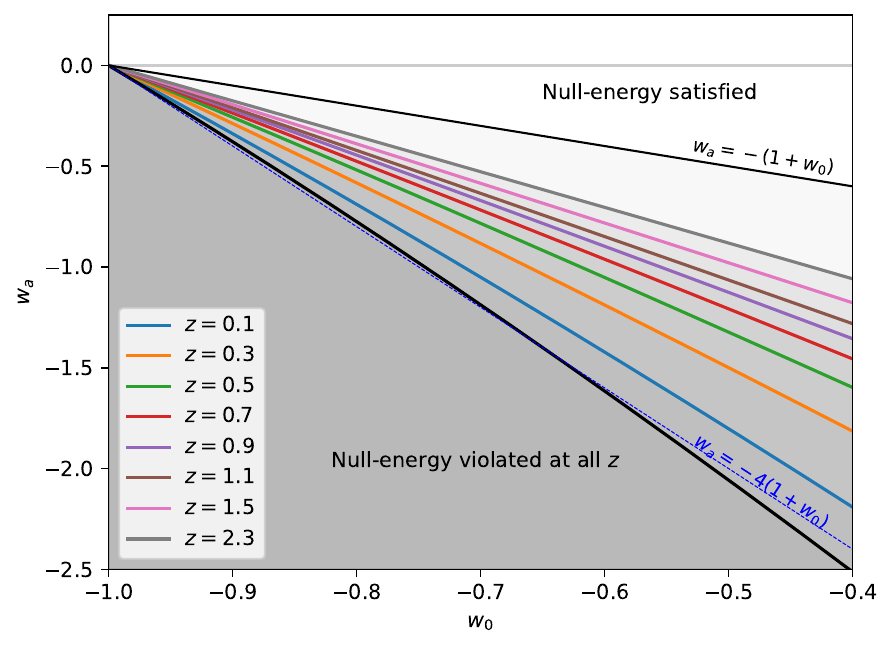}
    \caption{Regions of the $w_0$, $w_a$ plane where the sign of the predicted deviation of $\DM(z)$ from \LCDM\ is consistent with the null-energy condition,
      assuming given fixed $\theta_*$ and other non-dark energy parameters (taken to match \Planck). The lower diagonal of the plane cannot match the predicted sign of deviation at any redshift. The coloured lines show the lines above which the sign of the deviation is consistent for observations at various different redshifts. Above the $w_a=-(1+w_0)$ (upper black) line, the CPL model is consistent with the null energy condition, and hence predictions for observable deviations above this line have a consistent sign at all redshifts.
      The dashed blue line gives the approximation $w_a=-4(1+w_0)$ for the constant $H_0=H_0^\Lambda$ line (lower black) that divides the region where predictions are consistent at some redshift from where they are consistent at no redshift. All allowed regions also satisfy the $\DH/\DHL \le \DM/\DML$ bound.}
    \label{fig:w0waregions}
\end{figure}

For any point in the $w_0$, $w_a$ plane, a fixed value of $\theta_*$ and other non-dark energy parameters implies a specific value of the Hubble parameter today, $H_0$. We can therefore define a set of lines of constant $H_0$, one of which will coincide with the prediction of the \LCDM\ model at $w_0=-1$, $w_a=0$. For currently measured parameter values, this line approximately corresponds to $w_a \approx -4(1+w_0)$, though it is not exactly a straight line. For values of $w_a$ where  $w_a \alt -4(1+w_0)$, the implied Hubble parameter today is higher than in \LCDM, and for $w_a \agt -4(1+w_0)$ it is lower.  Since the null energy condition implies that $H_0$ can only decrease from the \LCDM\ model, this line divides the region where the sign of the deviation from \LCDM\ at $z \approx 0$ is consistent with the null-energy condition from the region where it is not.

BAO observations do not directly constrain $H_0$, since there are no galaxies at zero redshift. But for measurements at any given higher redshift, we can again ask a similar question: for which points in the $w_0$, $w_a$ plane does the CPL model predict a deviation from the \LCDM\ prediction for the observables that is consistent with the null-energy condition? For example, the null energy condition implies that $\DM\ge \DML$, so for observations at redshift $z$, there is a line in the $w_0$, $w_a$ plane where $\DM=\DML$ that divides regions where the predicted sign of the deviation is consistent with the null-energy condition, and where it is not. The $w_a \approx -4(1+w_0)$ line divides the regions where the observations are consistent at some redshift, from the region below the line where they are not consistent at any redshift. This is illustrated in Fig.~\ref{fig:w0waregions}. As also shown in Fig.~\ref{fig:BAO-null-regions}, for any point on the $w_a \approx -4(1+w_0)$ line, the predicted deviation of the observables from the \LCDM\ prediction becomes just inconsistent with the null-energy condition at all redshifts.

The significance of these separation lines for interpreting CPL-parameterization results is as follows: For data at redshift $z$, if the inferred $w_0$, $w_a$ constraint contour is below the corresponding line, the sign of the model fit relative to \LCDM\ is incorrect compared to any physical NEC-consistent model. Since \LCDM\ would give predictions exactly  on the line, the conclusion is that \LCDM\ would be preferred over any other null-energy consistent model. A detection of a signal clearly below the line would either indicate null-energy violation, some data or modelling error, or some more radical change in the cosmology (including changes from \LCDM\ around the time of recombination). 

For $w_0, w_a$ constraints on the NEC-consistent side of the \LCDM\ line, typically the closer the constraint is to the line, the less accurate the CPL-model prediction for that data point compared to a physical model. On the line itself, CPL predicts the same as \LCDM, so the error is 100\% of the size of the true model prediction deviation there. Constraints have to be a significant distance from the line for the CPL model predictions for the deviation to be accurate. As an example, Ref.~\cite{Shlivko:2024llw} consider a limiting thawing model with a sharp transition at low redshift, which gets close to the consistency line, but with deviations at the $0.7\%$ level in terms of observable predictions (comparable to the size of current data constraints).

The regions shown in Fig.~\ref{fig:w0waregions} are specifically for $\DM$, however the result for $\DV$ is qualitatively similar (with the intermediate redshift lines moving around slightly). For uncalibrated supernovae, the free overall normalization allows more freedom to match the behaviour of a null-energy consistent model, but the lower $w_a \approx -4(1+w_0)$ line corresponds roughly to the monotonic behaviour starting to be violated at around $z\approx 0.4$--$0.5$.

\begin{figure*}[th]
    \centering
    \includegraphics[width=\textwidth]{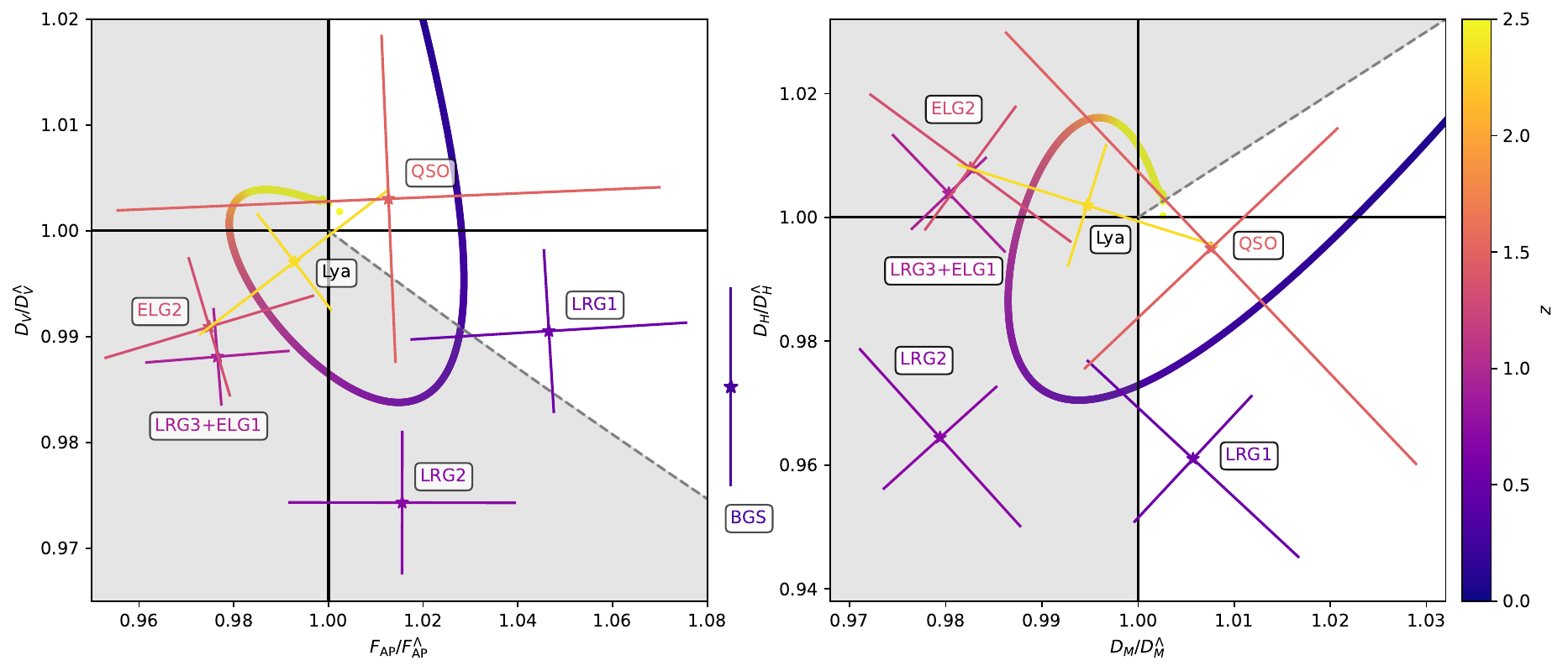}
    \caption{
         The null-energy exclusion regions assuming \Planck\ cosmological parameters compared to the DESI BAO data points~\citep{DESI:2024mwx}, similar to Fig.~\ref{fig:BAO-null-regions}. The error bars show the 1D marginalized standard deviations along the major and minor axes of the Gaussian error ellipses in plot coordinates (note the 2D $1\sigma$ contours lie $\sim 1.5\times$ further from the central point than the error bars, and the exact orientation of the bars depends on the aspect ratio of the plot). The coloured curves show the evolution of distances relative to \LCDM\ for the best fit DESI BAO + Early \Planck\ CPL model ($w_0=-0.50$, $w_a=-1.47$), with the colour bar now running to $z=2.5$. 
         The  BGS tracer has no reported value for $\DM/\rdrag$ or $\DH/\rdrag$, so that point is shown to the right of the left panel, with the star showing the corresponding value of $\DV/\DVL$ and error bars the standard deviation. 
         Parameter variations allow for some deviations from the fixed fiducial Planck 2018 model even in \LCDM, which is shown in Fig.~\ref{fig:param-depend} for comparison. 
     \label{fig:desi-BAO_null-regions}}
   
\end{figure*}

\section{DESI results}
\label{sec:DESI}

We now turn to considering the BAO results from the DESI Data Release 2 (Ref.~\citep{DESI:2025zgx}) in the context of the null energy condition. 
DESI provides a precision measurement of the BAO acoustic scale from various tracers at seven different effective redshifts. For six of these redshift bins DESI report a 2D covariance between $\DM/\rdrag$ and $\DH/\rdrag$, approximating the different redshifts as independent. For the other redshift there is less data, so they only report a single constraint on $\DV/\rdrag$ for that bin.

Fig.~\ref{fig:desi-BAO_null-regions} shows the DESI DR2 results compared to 
a fixed fiducial \LCDM\ model with \Planck\ PR3 best-fit cosmological parameters ($\theta_*$, $\Omega_m h^2$, and $\Omega_b h^2$~\cite{PL2018}). 
As in Fig.~\ref{fig:BAO-null-regions}, the null-energy exclusion regions summarized in Table~\ref{table:ineq} show the sections of the planes where deviation from \LCDM\ are allowed, assuming the fiducial non-dark energy parameters.
The right-hand panel shows the data points in the $\DH/\DHL$-$\DM/\DML$ plane, the left-hand panel shows the same data points mapped into the $\DV/\DVL$-$\FAP/\FAPL$ plane assuming Gaussian errors in $\DH$-$\DM$ with the reported correlation coefficient (plus the point where only $\DV/\rdrag$ was measured). Since the DESI full-shape analysis~\cite{DESI:2024hhd} does not qualitatively change the dark energy results, we focus here on the background geometric constraints from BAO alone.

It is visually clear from the plot in Fig.~\ref{fig:desi-BAO_null-regions} that if there is a deviation from the \LCDM\ prediction, it is mainly in the excluded region with $\DM <\DML$. For four of the redshift bins -- LRG2, LRG3+ELG1, ELG2, and Lya -- the data point centres lie within the exclusion region. Although the somewhat-outlying LRG1 point (effective redshift $z_{\rm eff}=0.51$) is in a notionally allowed region, this $\DV < \DVL$ section of the plane is hard to obtain in practice, since $\DV<\DVL$ is only obtainable at relatively high redshift (well beyond the \LCDM-equality redshift $z_c$, as explained in Sec.~\ref{sec:DV}). The QSO point lies outside the exclusion region and above $\DV > \DVL$, but it is important to note that the error bars are large and venture well into both the exclusion region and the allowed region where $\DV < \DVL$.

The coloured curves show the evolution of the best fit DESI BAO + Early \Planck\ flat CPL model with redshift, with $w_0=-0.50$ and $w_a = -1.47$.
Although the curves lie entirely within the exclusion region for redshifts $z \gtrsim 0.25$, the curve exits into the allowed region for present day. Since the CPL parameterization is not constrained by the null-energy condition, it can fit the data points slightly better by making compensating changes at low and high redshift ($w>-1$ at low redshift, phantom $w<-1$ at high redshift).

\begin{figure*}[tp]
    \centering
    \includegraphics[width=\columnwidth*2/3]{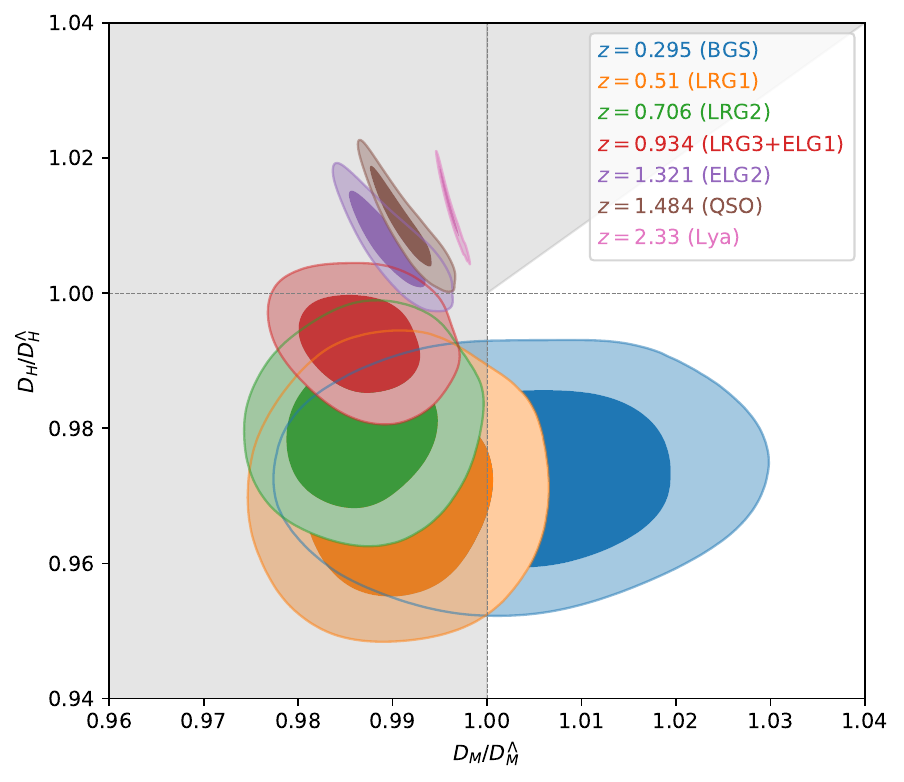}
    \includegraphics[width=\columnwidth*2/3]{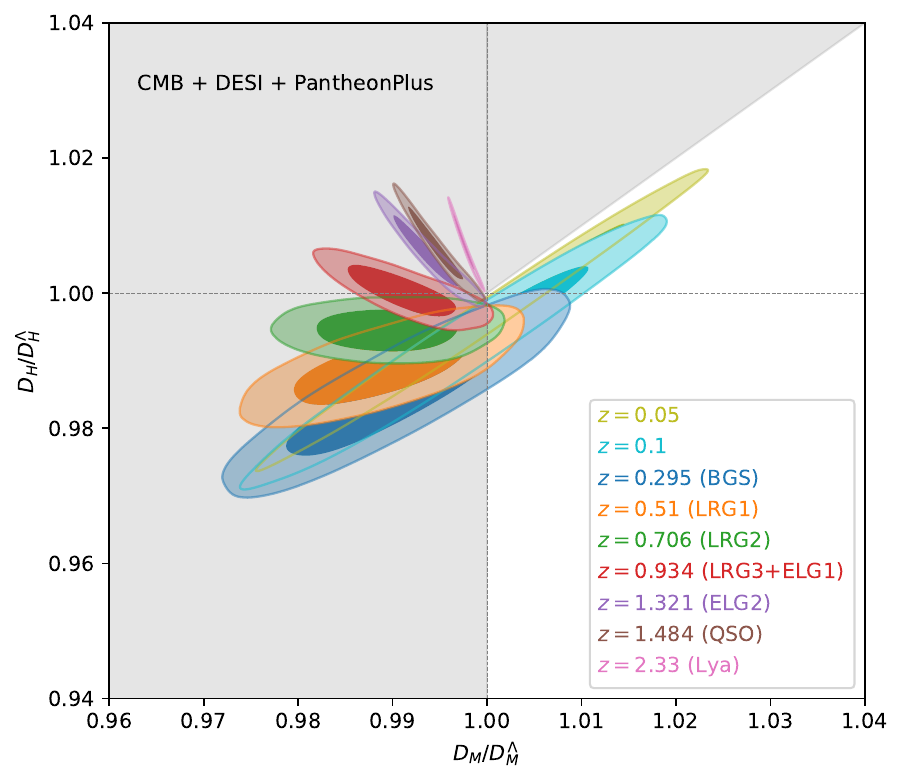}
    \includegraphics[width=\columnwidth*2/3]{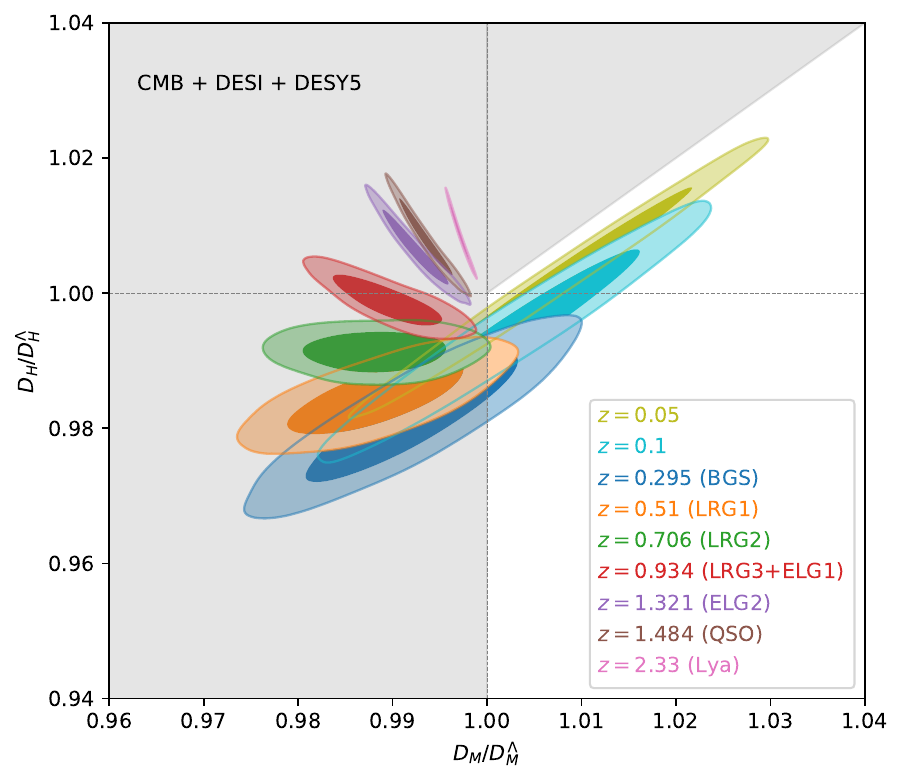}
   
    \caption{
         Posterior distributions of ratios of the DESI observables in the $w_0$, $w_a$ CPL model at various redshifts (colours, with corresponding nearest data point name in brackets in the legend), when combining DESI DR2 with \Planck\ constraints on $\theta_*$ and the early-\LCDM\ matter densities. 
         The left plot does not include supernovae data, the middle and right plot show the joint constraint including Pantheon Plus and DESY5 supernovae respectively.
         The $\DML$ and $\DHL$ values here are not from a fiducial model, but calculated as derived parameters for each point in the parameter chains (changing the dark energy model to $\Lambda$ at fixed $\theta_*$). Null-energy condition exclusion regions are shown in grey.
     \label{fig:DM-DH-posterior}}
   
\end{figure*}

We now consider how to interpret the DESI parameter constraints, focussing only on geometric constraints from the sound horizon on the background cosmological evolution.  We use the \Planck\ NPIPE (PR4~\citep{Akrami:2020bpw,Rosenberg:2022sdy}) data to constrain the high-redshift angular acoustic scale and matter densities.
Assuming standard \LCDM\ evolution until after recombination, \Planck\ constrains $\theta_*$, $\Omega_c h^2$,  $\Omega_b h^2$ to high precision, almost independently of the late-time dark energy or growth of structure~\cite{Lemos:2023xhs}.  We approximate the CMB likelihood in these three parameters using a Gaussian covariance calculated from the chains\footnote{\url{https://github.com/cmbant/PlanckEarlyLCDM}} of Ref.~\cite{Lemos:2023xhs}, and assume a minimal neutrino mass hierarchy approximated by one massive neutrino with $m_\nu=0.06{\rm eV}$. The DESI BAO likelihoods are calculated independently for each redshift, using the same \Cobaya\ likelihood code and approximations as Ref.~\cite{DESI:2024mwx,DESI:2025zgx}.
 We used the \Cobaya\ code \cite{Torrado:2020dgo} to run parameter chains from the likelihoods, and throughout we use $\CAMB$~\cite{Lewis:1999bs} to calculate the sound horizon, $\DM$, $\DH$ and other required quantities from a set of input cosmological parameters. \GetDist~\cite{Lewis:2019xzd} is used to calculate marginalized posteriors and plots from the parameter samples after removing burn in.

To illustrate how constraints can change when imposing the null-energy condition, we consider a modified NEC-consistent CPL model, implemented in a modified \CAMB\, to take 
\begin{equation}
    w(a)=\max{(w_0+w_a(1-a), -1)}.
    \label{eq:max}
\end{equation}
This agrees with CPL in the NEC-consistent region where $w_a>-(1+w_0)$, but modifies constraints below that.
We use wide prior ranges on the main cosmological parameters, and dark energy priors listed in the Appendix.~\ref{sec:priors} (Table~\ref{tab:params}). Note that for the max-$w_0w_a$CDM model of Eq.~\ref{eq:max}, the lower bound on the prior for $w_0$ was increased to $-1$, as for models with $w_a\leq0$, where the contours almost entirely lie, $w=-1$ for all $w_0<-1$. This stops the chains wandering around a region of equal likelihood.

Fig.~\ref{fig:DM-DH-posterior} shows predictions from the joint DESI (DR2, all redshifts) and CMB posterior, showing how the DESI observables evolve with redshift in the CPL posteriors. This can be compared with the data points in~Fig.~\ref{fig:desi-BAO_null-regions}.
 It is clear that the joint DESI posteriors do not favour NEC-consistent deviations from $\Lambda$ at any redshift, with most of the probability mass in the exclusion region with $\DM/\DML<1$ (and just over two-sigma away, though some of this is likely a parameter volume effect). 
 For any point in the posterior parameter space, after the first redshift bin the data would favour $\DM = \DML$ over the predictions of any other NEC-consistent model with $\DM > \DML$. The posteriors including supernovae show a similar pattern at DESI redshifts, while allowing the low-redshift ($z\alt 0.2$) evolution to have some NEC-consistent thawing-like behaviour that is slightly preferred by the uncalibrated supernova data. At DESI redshifts, the BAO data would however favour \LCDM\ over other NEC-consistent histories up to small statistical fluctuations (c.f. Ref.~\cite{DESI:2024kob}). For the supernovae, there are systematic differences between data sets due to differing samples and analysis choices as well as expected statistical fluctuations ~\cite{Gialamas:2024lyw,Efstathiou:2024xcq,Dhawan:2024gqy,Notari:2024zmi,DES:2025tir,Huang:2025som}. Any supernovae-driven pull towards NEC-consistent thawing would also make the distance ladder tension in the absolute calibration worse.

\begin{figure}[t!]
    \includegraphics[width=\columnwidth]{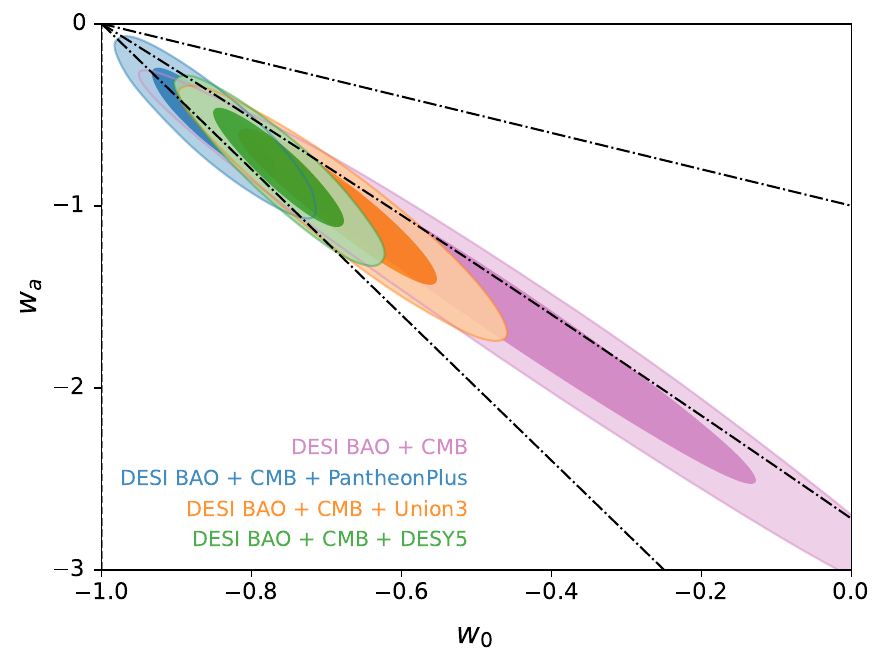}
    \caption{Posterior distributions of $w_0$ and $w_a$ for the CPL model from DESI BAO + CMB (pink), and DESI BAO + CMB + SN Ia as in figure 6 of Ref.~\cite{DESI:2024mwx}, using the \Planck\ early-\LCDM\ parameter constraints~\cite{Lemos:2023xhs}. Overplotted in the black dash-dotted lines are the bounds $w_a=-(1+w_0)$ and $w_a=-4(1+w_0)$, and the line in the $w_0w_a$ plane along which $\DM=\DML$ at $z=0.5$.}
    \label{fig:w0-wa-posterior}
\end{figure}

Fig.~\ref{fig:w0-wa-posterior} shows the joint CPL constraints in the $w_0$--$w_a$ plane. 
The joint constraints with uncalibrated supernovae are separately adding the same data constraints from Pantheon Plus~\cite{Brout:2022vxf}, Union~3~\cite{Rubin:2023ovl} and DES~Y5~\cite{DES:2024jxu} as in the DESI papers. With the addition of the supernova constraints, there is apparently some evidence for a deviation from \LCDM, though only marginally at the $2\sigma$-level in the case of Pantheon Plus.

However, the region of the $w_0$--$w_a$ plane that is favoured by the data is precisely the region where the CPL model does not predict the correct sign for BAO deviations from \LCDM\ with NEC-consistent models. The lines shown in Fig.~\ref{fig:w0-wa-posterior} correspond to the upper and lower null-energy consistency lines as in Fig.~\ref{fig:w0waregions}, while the middle line corresponds to where $\DM(z=0.5)=\DML(z=0.5)$. Below this middle line, the CPL model deviations from \LCDM\ are inconsistent with the sign of null-energy consistent deviations at all $z>0.5$ where the bulk of the DESI constraints lie. The constraint contours in the CPL plane therefore have no consistent interpretation in terms of additional information from BAO for constraints on NEC-consistent models.

\begin{figure}[t!]
    \includegraphics[width=\columnwidth]{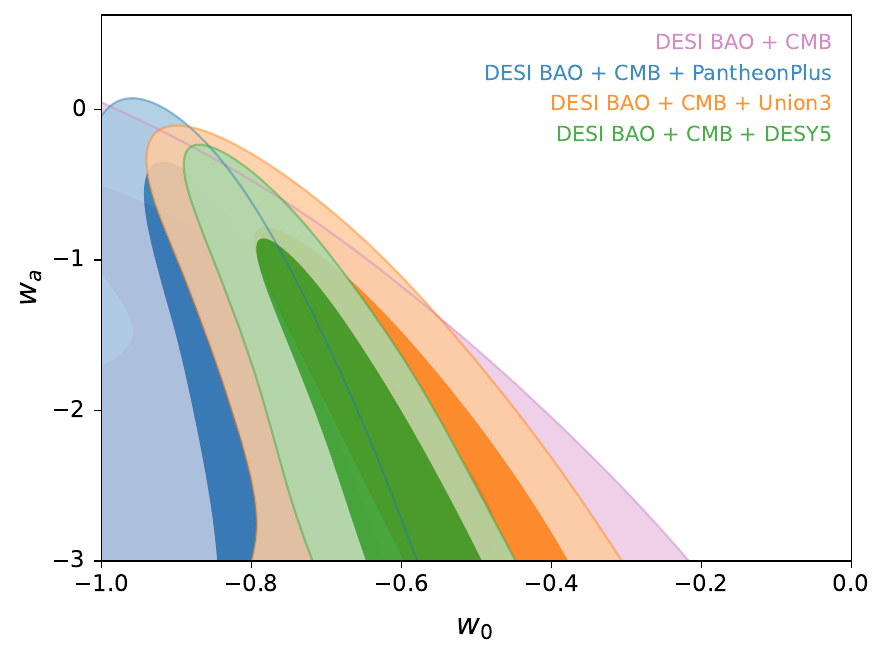}
    \caption{Posterior distributions of $w_0$ and $w_a$ as in Fig. \ref{fig:w0-wa-posterior}, except now using the $w(a)=\max{(w_0+w_a(1-a), -1)}$ model that imposes NEC-consistent evolution. For reference, the dashed line shows the Pantheon Plus constraint from Fig. \ref{fig:w0-wa-posterior} (the normal CPL parameterization where NEC can be violated).}
    \label{fig:max-w0-wa-posterior}
\end{figure}

In Fig.~\ref{fig:max-w0-wa-posterior} we show the constraints in the $w_0$--$w_a$ plane for the Max-$w_0w_a$CDM model of Eq.~\ref{eq:max}. There is now a degeneracy near the \LCDM\ line at $w_0= -1$ (for $w_a\le 0$) where the model is nearly independent of $w_a$, so marginalized constraints depend strongly on the prior ranges. However, we can see that the joint constraints with Pantheon Plus are now more consistent with \LCDM, with the deviations from \LCDM\ being driven by the other supernova data, esp. DES Y5,  pulling towards thawing-like behaviour at low redshift.
 While this is just a simple non-physical toy model, since $w(a)\geq-1$ everywhere, this model is sufficient to illustrate how the constraints and degeneracy directions depend on the NEC-violating part of the CPL parameter space.

\section{Conclusions}

The cosmic microwave background provides a robust measurement of the matter densities and angular acoustic scale $\theta_*$, which strongly constrains possible deviations from \LCDM\ when combined with lower-redshift baryon acoustic oscillation (BAO) measurements. We have shown that the null energy condition, which requires that a non-interactive dark energy must have an equation of state parameter $w\ge -1$, imposes strict inequalities on BAO observables relative to \LCDM\ predictions. These constraints rule out substantial regions of parameter space for any physical dark energy model that respects this condition. Although there is no clear reason that the null-energy condition cannot be violated at the level of the background cosmology~\cite{Kaplinghat:2003vf}, a violation would indicate some qualitatively new physics compared to simple models where the null-energy condition always holds (such as quintessence scalar fields with standard kinetic terms).

Current DESI BAO measurements exhibit small tensions with \LCDM, with a couple of data points deviating by approximately 2$\sigma$ in directions that cannot be explained by any NEC-consistent dark energy model. 
While the CPL parameterization $w(a) = w_0 + w_a(1-a)$ can fit these deviations to some extent, the model requires phantom-like behaviour at intermediate redshift that violates the null energy condition when interpreted as a dark energy model. This highlights the importance of physical constraints when interpreting apparent deviations from \LCDM\ using simple parameterizations. Current acoustic data favour \LCDM\ over other NEC-consistent models up to statistical fluctuations. 
While our broad conclusions are not new~\cite{Linder:2024rdj,DESI:2024aqx,DESI:2025zgx}, and some of our individual results are well known, our analysis provides new tools for interpreting BAO data more directly through the lens of physical constraints.

While we have focused here on geometric constraints from the acoustic standard ruler, a complete picture will require combining these results with constraints from structure growth and other cosmological probes. Although the CPL parameterization is convenient and easy to implement, it is fundamentally unphysical, so results using it must be interpreted with care. Future work should explore how to robustly identify physically-consistent deviations from \LCDM\ when combining with other observables.  
If future data shows tensions in directions forbidden by the null energy condition, this would point toward either new systematic effects in the measurements or the need for more radical modifications to our cosmological model. Such modifications could include interacting dark energy, modifications to gravity, or changes to early-Universe physics. 
However, the one-sided nature of the acoustic scale constraints show that current acoustic scale data are consistent with $\Lambda$CDM or tend to pull towards the excluded regions, not favouring deviations from $\Lambda$CDM that are consistent with the null-energy condition. Any significant preference for NEC-consistent deviations from \planck\ $\Lambda$CDM~\cite{DES:2024jxu,Rubin:2023ovl,DESI:2024kob,Shajib:2025tpd,DESI:2025zgx} is primarily driven by supernovae and depends somewhat on the dataset used.

\begin{acknowledgments}
AL is supported by the UK STFC grant ST/X001040/1. EC is supported by a UK Science and Technology Facilities Council (STFC) studentship. We thank Robert Scherrer, Willian Wolf and Sesh Nadathur for comments and questions.
\end{acknowledgments}

\appendix

\section{Monotonicity of \( R(z) = \DH / \DHL \) at $z\le z_c$}
\label{appendix:proof}

To establish that \( R(z) \) is a decreasing function for \( z \leq z_c \), we compute its derivative with respect to \( z \):
\begin{equation}
\frac{dR}{dz} = \frac{d}{dz} \left( \frac{\DH}{\DHL} \right) = \frac{\DHL \, \frac{d\DH}{dz} - \DH \, \frac{d\DHL}{dz}}{\DHL{}^2}.
\end{equation}

From Eq.~\eqref{eq:hubblederiv}, we have
\begin{equation}
\frac{d\DH}{dz} \leq \left( \frac{\DH}{\DHL} \right)^3 \frac{d\DHL}{dz} = R(z)^3 \frac{d\DHL}{dz}.
\end{equation}

Substituting this inequality into the expression for \( \frac{dR}{dz} \), we obtain
\begin{align}
\frac{dR}{dz} &\leq \frac{ \DHL \left( R(z)^3 \frac{d\DHL}{dz} \right) - \DH \frac{d\DHL}{dz} }{ \DHL{}^2 } \nonumber \\
&= \frac{ \left( R(z)^3 \DHL - R(z) \DHL \right) }{ \DHL{}^2 } \frac{d\DHL}{dz} \nonumber \\
&= \frac{ R(z) \left( R(z)^2 - 1 \right) }{ \DHL } \frac{d\DHL}{dz}.
\end{align}

For \( z \leq z_c \), we have \( R(z) > 1 \) and \( \frac{d\DHL}{dz} < 0 \) because \( \DHL(z) \) is monotonically decreasing. Therefore,
\begin{equation}
\frac{dR}{dz} \leq \frac{ R(z) \left( R(z)^2 - 1 \right) }{ \DHL } \left( \frac{d\DHL}{dz} \right) < 0.
\end{equation}
This shows that \( R(z) \) is monotonically decreasing for \( z \leq z_c \).

\section{Priors}
\label{sec:priors}

\begin{table}[h!]
\centering
\begin{tabular}{ c c c }
    \hline
    \textbf{Model} & \textbf{Parameter} & \textbf{Prior} \\
    \hline
    $w_0w_a$CDM & $w_0$ & $\mathcal U[-3, 0]$ \\
        & $w_a$ & $\mathcal U[-3, 2]$ \\
    \hline
    Max-$w_0w_a$CDM & $w_0$ & $\mathcal U[-1, 0]$ \\
        & $w_a$ & $\mathcal U[-3, 2]$ \\ 
    \hline
\end{tabular}
\caption{Priors on $w_0$ and $w_a$ used in the analysis of the DESI BAO + CMB and DESI BAO + CMB + SN Ia data sets. The Max-$w_0w_a$CDM model is the model of Eq.~\eqref{eq:max}.}
\label{tab:params}
\end{table}

\vspace{1cm}

\widetext
\quad\\
\FloatBarrier  
\begin{figure*}[!h]
    \centering
    \includegraphics[width=0.4\columnwidth]{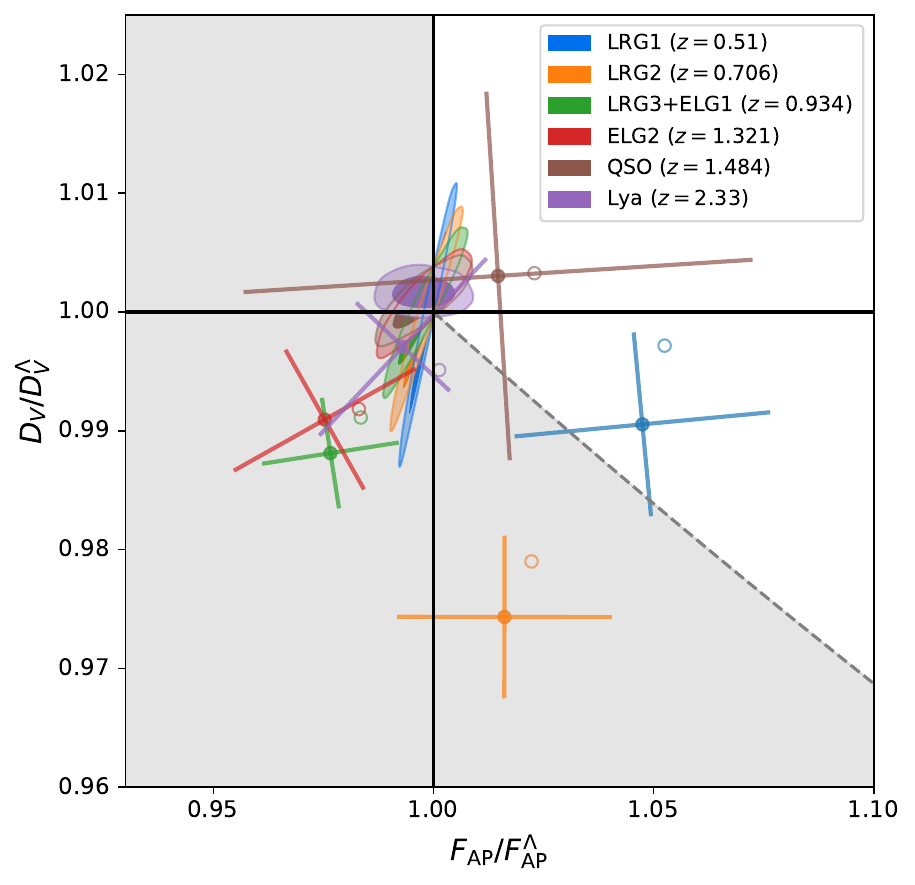}
    \includegraphics[width=0.4\columnwidth]{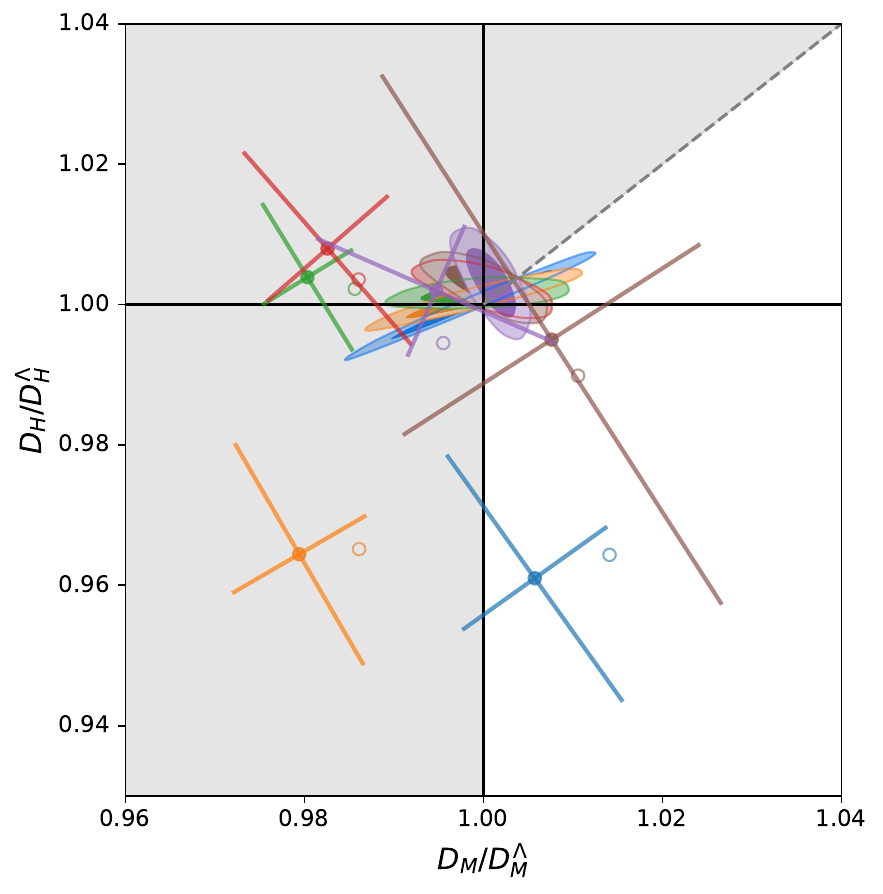}    
    \caption{Similar to Fig.~\ref{fig:desi-BAO_null-regions}, showing deviations with respect to a fixed best-fit \Planck\ model for the DESI DR2 data points, but also showing the variation with redshift of the \planck\ early-$\Lambda$CDM posterior~\cite{Lemos:2023xhs} in these parameters (contours). This illustrates how much predictions of the observables can change only due to shifts in $\Lambda$CDM cosmological parameters allowed by the \planck\ data. The contours are slightly off centre due to small parameter shifts in the more recent  \planck\ early-$\Lambda$CDM likelihood and differences due to this likelihood not assuming $\Lambda$CDM lensing at late times.
    The hollow points show where the DESI data points shift to if the reference \LCDM\ model is taken to be a joint
    best fit from the combination of \planck\ NPIPE and DESI DR2 (which pushes parameter constraints into a corner of the \planck\ posterior, with somewhat lower $\Omega_m$ [$\Omega_m\approx 0.302$] compared to the \Planck-only 2018 fit).
    Note that Fig.~\ref{fig:DM-DH-posterior} already accounts for parameter variations, since that figure shows ratios evaluated at each point in non-dark-energy parameter space rather than relative to a fixed fiducial model. 
    \label{fig:param-depend}
    }
\end{figure*}
\clearpage
\vspace{0pt}\noindent\endwidetext
\providecommand{\aj}{Astron. J. }\providecommand{\apj}{ApJ
  }\providecommand{\apjl}{ApJ
  }\providecommand{\mnras}{MNRAS}\providecommand{\prl}{PRL}\providecommand{\prd}{PRD}\providecommand{\jcap}{JCAP}\providecommand{\aap}{A\&A}


\begin{thebibliography}{41}%
\makeatletter
\providecommand \@ifxundefined [1]{%
 \@ifx{#1\undefined}
}%
\providecommand \@ifnum [1]{%
 \ifnum #1\expandafter \@firstoftwo
 \else \expandafter \@secondoftwo
 \fi
}%
\providecommand \@ifx [1]{%
 \ifx #1\expandafter \@firstoftwo
 \else \expandafter \@secondoftwo
 \fi
}%
\providecommand \natexlab [1]{#1}%
\providecommand \enquote  [1]{``#1''}%
\providecommand \bibnamefont  [1]{#1}%
\providecommand \bibfnamefont [1]{#1}%
\providecommand \citenamefont [1]{#1}%
\providecommand \href@noop [0]{\@secondoftwo}%
\providecommand \href [0]{\begingroup \@sanitize@url \@href}%
\providecommand \@href[1]{\@@startlink{#1}\@@href}%
\providecommand \@@href[1]{\endgroup#1\@@endlink}%
\providecommand \@sanitize@url [0]{\catcode `\\12\catcode `\$12\catcode
  `\&12\catcode `\#12\catcode `\^12\catcode `\_12\catcode `\%12\relax}%
\providecommand \@@startlink[1]{}%
\providecommand \@@endlink[0]{}%
\providecommand \url  [0]{\begingroup\@sanitize@url \@url }%
\providecommand \@url [1]{\endgroup\@href {#1}{\urlprefix }}%
\providecommand \urlprefix  [0]{URL }%
\providecommand \Eprint [0]{\href }%
\providecommand \doibase [0]{https://doi.org/}%
\providecommand \selectlanguage [0]{\@gobble}%
\providecommand \bibinfo  [0]{\@secondoftwo}%
\providecommand \bibfield  [0]{\@secondoftwo}%
\providecommand \translation [1]{[#1]}%
\providecommand \BibitemOpen [0]{}%
\providecommand \bibitemStop [0]{}%
\providecommand \bibitemNoStop [0]{.\EOS\space}%
\providecommand \EOS [0]{\spacefactor3000\relax}%
\providecommand \BibitemShut  [1]{\csname bibitem#1\endcsname}%
\let\auto@bib@innerbib\@empty
\bibitem [{\citenamefont {Aghanim}\ \emph
  {et~al.}(2020{\natexlab{a}})\citenamefont {Aghanim} \emph
  {et~al.}}]{PCP2018}%
  \BibitemOpen
  \bibfield  {author} {\bibinfo {author} {\bibfnamefont {N.}~\bibnamefont
  {Aghanim}} \emph {et~al.} (\bibinfo {collaboration} {Planck}),\ }\bibfield
  {title} {\bibinfo {title} {{Planck 2018 results. VI. Cosmological
  parameters}},\ }\href {https://doi.org/10.1051/0004-6361/201833910}
  {\bibfield  {journal} {\bibinfo  {journal} {\aap}\ }\textbf {\bibinfo
  {volume} {641}},\ \bibinfo {pages} {A6} (\bibinfo {year}
  {2020}{\natexlab{a}})},\ \Eprint {https://arxiv.org/abs/1807.06209}
  {arXiv:1807.06209 [astro-ph.CO]} \BibitemShut {NoStop}%
\bibitem [{\citenamefont {Lemos}\ and\ \citenamefont
  {Lewis}(2023)}]{Lemos:2023xhs}%
  \BibitemOpen
  \bibfield  {author} {\bibinfo {author} {\bibfnamefont {P.}~\bibnamefont
  {Lemos}}\ and\ \bibinfo {author} {\bibfnamefont {A.}~\bibnamefont {Lewis}},\
  }\bibfield  {title} {\bibinfo {title} {{CMB constraints on the early Universe
  independent of late-time cosmology}},\ }\href
  {https://doi.org/10.1103/PhysRevD.107.103505} {\bibfield  {journal} {\bibinfo
   {journal} {Phys. Rev. D}\ }\textbf {\bibinfo {volume} {107}},\ \bibinfo
  {pages} {103505} (\bibinfo {year} {2023})},\ \Eprint
  {https://arxiv.org/abs/2302.12911} {arXiv:2302.12911 [astro-ph.CO]}
  \BibitemShut {NoStop}%
\bibitem [{\citenamefont {Adame}\ \emph {et~al.}(2025)\citenamefont {Adame}
  \emph {et~al.}}]{DESI:2024mwx}%
  \BibitemOpen
  \bibfield  {author} {\bibinfo {author} {\bibfnamefont {A.~G.}\ \bibnamefont
  {Adame}} \emph {et~al.} (\bibinfo {collaboration} {DESI}),\ }\bibfield
  {title} {\bibinfo {title} {{DESI 2024 VI: cosmological constraints from the
  measurements of baryon acoustic oscillations}},\ }\href
  {https://doi.org/10.1088/1475-7516/2025/02/021} {\bibfield  {journal}
  {\bibinfo  {journal} {JCAP}\ }\textbf {\bibinfo {volume} {02}},\ \bibinfo
  {pages} {021}},\ \Eprint {https://arxiv.org/abs/2404.03002} {arXiv:2404.03002
  [astro-ph.CO]} \BibitemShut {NoStop}%
\bibitem [{\citenamefont {Abdul~Karim}\ \emph {et~al.}(2025)\citenamefont
  {Abdul~Karim} \emph {et~al.}}]{DESI:2025zgx}%
  \BibitemOpen
  \bibfield  {author} {\bibinfo {author} {\bibfnamefont {M.}~\bibnamefont
  {Abdul~Karim}} \emph {et~al.} (\bibinfo {collaboration} {DESI}),\ }\bibfield
  {title} {\bibinfo {title} {{DESI DR2 Results II: Measurements of Baryon
  Acoustic Oscillations and Cosmological Constraints}},\ }\href@noop {} {\
  (\bibinfo {year} {2025})},\ \Eprint {https://arxiv.org/abs/2503.14738}
  {arXiv:2503.14738 [astro-ph.CO]} \BibitemShut {NoStop}%
\bibitem [{\citenamefont {Calderon}\ \emph {et~al.}(2024)\citenamefont
  {Calderon} \emph {et~al.}}]{DESI:2024aqx}%
  \BibitemOpen
  \bibfield  {author} {\bibinfo {author} {\bibfnamefont {R.}~\bibnamefont
  {Calderon}} \emph {et~al.} (\bibinfo {collaboration} {DESI}),\ }\bibfield
  {title} {\bibinfo {title} {{DESI 2024: reconstructing dark energy using
  crossing statistics with DESI DR1 BAO data}},\ }\href
  {https://doi.org/10.1088/1475-7516/2024/10/048} {\bibfield  {journal}
  {\bibinfo  {journal} {JCAP}\ }\textbf {\bibinfo {volume} {10}},\ \bibinfo
  {pages} {048}},\ \Eprint {https://arxiv.org/abs/2405.04216} {arXiv:2405.04216
  [astro-ph.CO]} \BibitemShut {NoStop}%
\bibitem [{\citenamefont {Lodha}\ \emph
  {et~al.}(2025{\natexlab{a}})\citenamefont {Lodha} \emph
  {et~al.}}]{DESI:2024kob}%
  \BibitemOpen
  \bibfield  {author} {\bibinfo {author} {\bibfnamefont {K.}~\bibnamefont
  {Lodha}} \emph {et~al.} (\bibinfo {collaboration} {DESI}),\ }\bibfield
  {title} {\bibinfo {title} {{DESI 2024: Constraints on physics-focused aspects
  of dark energy using DESI DR1 BAO data}},\ }\href
  {https://doi.org/10.1103/PhysRevD.111.023532} {\bibfield  {journal} {\bibinfo
   {journal} {Phys. Rev. D}\ }\textbf {\bibinfo {volume} {111}},\ \bibinfo
  {pages} {023532} (\bibinfo {year} {2025}{\natexlab{a}})},\ \Eprint
  {https://arxiv.org/abs/2405.13588} {arXiv:2405.13588 [astro-ph.CO]}
  \BibitemShut {NoStop}%
\bibitem [{\citenamefont {Cort\^es}\ and\ \citenamefont
  {Liddle}(2024)}]{Cortes:2024lgw}%
  \BibitemOpen
  \bibfield  {author} {\bibinfo {author} {\bibfnamefont {M.}~\bibnamefont
  {Cort\^es}}\ and\ \bibinfo {author} {\bibfnamefont {A.~R.}\ \bibnamefont
  {Liddle}},\ }\bibfield  {title} {\bibinfo {title} {{Interpreting DESI's
  evidence for evolving dark energy}},\ }\href
  {https://doi.org/10.1088/1475-7516/2024/12/007} {\bibfield  {journal}
  {\bibinfo  {journal} {JCAP}\ }\textbf {\bibinfo {volume} {12}},\ \bibinfo
  {pages} {007}},\ \Eprint {https://arxiv.org/abs/2404.08056} {arXiv:2404.08056
  [astro-ph.CO]} \BibitemShut {NoStop}%
\bibitem [{\citenamefont {Linder}(2024)}]{Linder:2024rdj}%
  \BibitemOpen
  \bibfield  {author} {\bibinfo {author} {\bibfnamefont {E.~V.}\ \bibnamefont
  {Linder}},\ }\bibfield  {title} {\bibinfo {title} {{Interpreting Dark Energy
  Data Away from $\Lambda$}},\ }\href@noop {} {\  (\bibinfo {year} {2024})},\
  \Eprint {https://arxiv.org/abs/2410.10981} {arXiv:2410.10981 [astro-ph.CO]}
  \BibitemShut {NoStop}%
\bibitem [{\citenamefont {Fikri}\ \emph {et~al.}(2024)\citenamefont {Fikri},
  \citenamefont {ElKhateeb}, \citenamefont {Lashin},\ and\ \citenamefont
  {El~Hanafy}}]{Fikri:2024klc}%
  \BibitemOpen
  \bibfield  {author} {\bibinfo {author} {\bibfnamefont {R.}~\bibnamefont
  {Fikri}}, \bibinfo {author} {\bibfnamefont {E.}~\bibnamefont {ElKhateeb}},
  \bibinfo {author} {\bibfnamefont {E.~S.}\ \bibnamefont {Lashin}},\ and\
  \bibinfo {author} {\bibfnamefont {W.}~\bibnamefont {El~Hanafy}},\ }\bibfield
  {title} {\bibinfo {title} {{A preference for dynamical phantom dark energy
  using one-parameter model with Planck, DESI DR1 BAO and SN data}},\
  }\href@noop {} {\  (\bibinfo {year} {2024})},\ \Eprint
  {https://arxiv.org/abs/2411.19362} {arXiv:2411.19362 [astro-ph.CO]}
  \BibitemShut {NoStop}%
\bibitem [{\citenamefont {Lodha}\ \emph
  {et~al.}(2025{\natexlab{b}})\citenamefont {Lodha} \emph
  {et~al.}}]{DESI:2025fii}%
  \BibitemOpen
  \bibfield  {author} {\bibinfo {author} {\bibfnamefont {K.}~\bibnamefont
  {Lodha}} \emph {et~al.} (\bibinfo {collaboration} {DESI}),\ }\bibfield
  {title} {\bibinfo {title} {{Extended Dark Energy analysis using DESI DR2 BAO
  measurements}},\ }\href@noop {} {\  (\bibinfo {year} {2025}{\natexlab{b}})},\
  \Eprint {https://arxiv.org/abs/2503.14743} {arXiv:2503.14743 [astro-ph.CO]}
  \BibitemShut {NoStop}%
\bibitem [{\citenamefont {Sen}\ and\ \citenamefont
  {Scherrer}(2008)}]{Sen:2007ep}%
  \BibitemOpen
  \bibfield  {author} {\bibinfo {author} {\bibfnamefont {A.~A.}\ \bibnamefont
  {Sen}}\ and\ \bibinfo {author} {\bibfnamefont {R.~J.}\ \bibnamefont
  {Scherrer}},\ }\bibfield  {title} {\bibinfo {title} {{The Weak Energy
  Condition and the Expansion History of the Universe}},\ }\href
  {https://doi.org/10.1016/j.physletb.2007.11.070} {\bibfield  {journal}
  {\bibinfo  {journal} {Phys. Lett. B}\ }\textbf {\bibinfo {volume} {659}},\
  \bibinfo {pages} {457} (\bibinfo {year} {2008})},\ \Eprint
  {https://arxiv.org/abs/astro-ph/0703416} {arXiv:astro-ph/0703416}
  \BibitemShut {NoStop}%
\bibitem [{\citenamefont {Chevallier}\ and\ \citenamefont
  {Polarski}(2001)}]{Chevallier:2000qy}%
  \BibitemOpen
  \bibfield  {author} {\bibinfo {author} {\bibfnamefont {M.}~\bibnamefont
  {Chevallier}}\ and\ \bibinfo {author} {\bibfnamefont {D.}~\bibnamefont
  {Polarski}},\ }\bibfield  {title} {\bibinfo {title} {{Accelerating universes
  with scaling dark matter}},\ }\href
  {https://doi.org/10.1142/S0218271801000822} {\bibfield  {journal} {\bibinfo
  {journal} {Int. J. Mod. Phys. D}\ }\textbf {\bibinfo {volume} {10}},\
  \bibinfo {pages} {213} (\bibinfo {year} {2001})},\ \Eprint
  {https://arxiv.org/abs/gr-qc/0009008} {arXiv:gr-qc/0009008} \BibitemShut
  {NoStop}%
\bibitem [{\citenamefont {Linder}(2003)}]{Linder:2002et}%
  \BibitemOpen
  \bibfield  {author} {\bibinfo {author} {\bibfnamefont {E.~V.}\ \bibnamefont
  {Linder}},\ }\bibfield  {title} {\bibinfo {title} {{Exploring the expansion
  history of the universe}},\ }\href
  {https://doi.org/10.1103/PhysRevLett.90.091301} {\bibfield  {journal}
  {\bibinfo  {journal} {Phys. Rev. Lett.}\ }\textbf {\bibinfo {volume} {90}},\
  \bibinfo {pages} {091301} (\bibinfo {year} {2003})},\ \Eprint
  {https://arxiv.org/abs/astro-ph/0208512} {arXiv:astro-ph/0208512}
  \BibitemShut {NoStop}%
\bibitem [{\citenamefont {Shlivko}\ and\ \citenamefont
  {Steinhardt}(2024)}]{Shlivko:2024llw}%
  \BibitemOpen
  \bibfield  {author} {\bibinfo {author} {\bibfnamefont {D.}~\bibnamefont
  {Shlivko}}\ and\ \bibinfo {author} {\bibfnamefont {P.~J.}\ \bibnamefont
  {Steinhardt}},\ }\bibfield  {title} {\bibinfo {title} {{Assessing
  observational constraints on dark energy}},\ }\href
  {https://doi.org/10.1016/j.physletb.2024.138826} {\bibfield  {journal}
  {\bibinfo  {journal} {Phys. Lett. B}\ }\textbf {\bibinfo {volume} {855}},\
  \bibinfo {pages} {138826} (\bibinfo {year} {2024})},\ \Eprint
  {https://arxiv.org/abs/2405.03933} {arXiv:2405.03933 [astro-ph.CO]}
  \BibitemShut {NoStop}%
\bibitem [{\citenamefont {Wolf}\ \emph {et~al.}(2024)\citenamefont {Wolf},
  \citenamefont {Garc\'\i{}a-Garc\'\i{}a}, \citenamefont {Bartlett},\ and\
  \citenamefont {Ferreira}}]{Wolf:2024eph}%
  \BibitemOpen
  \bibfield  {author} {\bibinfo {author} {\bibfnamefont {W.~J.}\ \bibnamefont
  {Wolf}}, \bibinfo {author} {\bibfnamefont {C.}~\bibnamefont
  {Garc\'\i{}a-Garc\'\i{}a}}, \bibinfo {author} {\bibfnamefont {D.~J.}\
  \bibnamefont {Bartlett}},\ and\ \bibinfo {author} {\bibfnamefont {P.~G.}\
  \bibnamefont {Ferreira}},\ }\bibfield  {title} {\bibinfo {title} {{Scant
  evidence for thawing quintessence}},\ }\href
  {https://doi.org/10.1103/PhysRevD.110.083528} {\bibfield  {journal} {\bibinfo
   {journal} {Phys. Rev. D}\ }\textbf {\bibinfo {volume} {110}},\ \bibinfo
  {pages} {083528} (\bibinfo {year} {2024})},\ \Eprint
  {https://arxiv.org/abs/2408.17318} {arXiv:2408.17318 [astro-ph.CO]}
  \BibitemShut {NoStop}%
\bibitem [{\citenamefont {Payeur}\ \emph {et~al.}(2024)\citenamefont {Payeur},
  \citenamefont {McDonough},\ and\ \citenamefont
  {Brandenberger}}]{Payeur:2024dnq}%
  \BibitemOpen
  \bibfield  {author} {\bibinfo {author} {\bibfnamefont {G.}~\bibnamefont
  {Payeur}}, \bibinfo {author} {\bibfnamefont {E.}~\bibnamefont {McDonough}},\
  and\ \bibinfo {author} {\bibfnamefont {R.}~\bibnamefont {Brandenberger}},\
  }\bibfield  {title} {\bibinfo {title} {{Do Observations Prefer Thawing
  Quintessence?}},\ }\href@noop {} {\  (\bibinfo {year} {2024})},\ \Eprint
  {https://arxiv.org/abs/2411.13637} {arXiv:2411.13637 [astro-ph.CO]}
  \BibitemShut {NoStop}%
\bibitem [{\citenamefont {Riess}\ \emph {et~al.}(2022)\citenamefont {Riess},
  \citenamefont {Breuval}, \citenamefont {Yuan}, \citenamefont {Casertano},
  \citenamefont {Macri}, \citenamefont {Bowers}, \citenamefont {Scolnic},
  \citenamefont {Cantat-Gaudin}, \citenamefont {Anderson},\ and\ \citenamefont
  {Reyes}}]{Riess:2022mme}%
  \BibitemOpen
  \bibfield  {author} {\bibinfo {author} {\bibfnamefont {A.~G.}\ \bibnamefont
  {Riess}}, \bibinfo {author} {\bibfnamefont {L.}~\bibnamefont {Breuval}},
  \bibinfo {author} {\bibfnamefont {W.}~\bibnamefont {Yuan}}, \bibinfo {author}
  {\bibfnamefont {S.}~\bibnamefont {Casertano}}, \bibinfo {author}
  {\bibfnamefont {L.~M.}\ \bibnamefont {Macri}}, \bibinfo {author}
  {\bibfnamefont {J.~B.}\ \bibnamefont {Bowers}}, \bibinfo {author}
  {\bibfnamefont {D.}~\bibnamefont {Scolnic}}, \bibinfo {author} {\bibfnamefont
  {T.}~\bibnamefont {Cantat-Gaudin}}, \bibinfo {author} {\bibfnamefont {R.~I.}\
  \bibnamefont {Anderson}},\ and\ \bibinfo {author} {\bibfnamefont {M.~C.}\
  \bibnamefont {Reyes}},\ }\bibfield  {title} {\bibinfo {title} {{Cluster
  Cepheids with High Precision Gaia Parallaxes, Low Zero-point Uncertainties,
  and Hubble Space Telescope Photometry}},\ }\href
  {https://doi.org/10.3847/1538-4357/ac8f24} {\bibfield  {journal} {\bibinfo
  {journal} {\apj}\ }\textbf {\bibinfo {volume} {938}},\ \bibinfo {pages} {36}
  (\bibinfo {year} {2022})},\ \Eprint {https://arxiv.org/abs/2208.01045}
  {arXiv:2208.01045 [astro-ph.CO]} \BibitemShut {NoStop}%
\bibitem [{\citenamefont {Freedman}\ \emph {et~al.}(2024)\citenamefont
  {Freedman}, \citenamefont {Madore}, \citenamefont {Jang}, \citenamefont
  {Hoyt}, \citenamefont {Lee},\ and\ \citenamefont {Owens}}]{Freedman:2024eph}%
  \BibitemOpen
  \bibfield  {author} {\bibinfo {author} {\bibfnamefont {W.~L.}\ \bibnamefont
  {Freedman}}, \bibinfo {author} {\bibfnamefont {B.~F.}\ \bibnamefont
  {Madore}}, \bibinfo {author} {\bibfnamefont {I.~S.}\ \bibnamefont {Jang}},
  \bibinfo {author} {\bibfnamefont {T.~J.}\ \bibnamefont {Hoyt}}, \bibinfo
  {author} {\bibfnamefont {A.~J.}\ \bibnamefont {Lee}},\ and\ \bibinfo {author}
  {\bibfnamefont {K.~A.}\ \bibnamefont {Owens}},\ }\bibfield  {title} {\bibinfo
  {title} {{Status Report on the Chicago-Carnegie Hubble Program (CCHP): Three
  Independent Astrophysical Determinations of the Hubble Constant Using the
  James Webb Space Telescope}},\ }\href@noop {} {\  (\bibinfo {year} {2024})},\
  \Eprint {https://arxiv.org/abs/2408.06153} {arXiv:2408.06153 [astro-ph.CO]}
  \BibitemShut {NoStop}%
\bibitem [{\citenamefont {Riess}\ \emph {et~al.}(2024)\citenamefont {Riess}
  \emph {et~al.}}]{Riess:2024vfa}%
  \BibitemOpen
  \bibfield  {author} {\bibinfo {author} {\bibfnamefont {A.~G.}\ \bibnamefont
  {Riess}} \emph {et~al.},\ }\bibfield  {title} {\bibinfo {title} {{JWST
  Validates HST Distance Measurements: Selection of Supernova Subsample
  Explains Differences in JWST Estimates of Local H $_{0}$}},\ }\href
  {https://doi.org/10.3847/1538-4357/ad8c21} {\bibfield  {journal} {\bibinfo
  {journal} {Astrophys. J.}\ }\textbf {\bibinfo {volume} {977}},\ \bibinfo
  {pages} {120} (\bibinfo {year} {2024})},\ \Eprint
  {https://arxiv.org/abs/2408.11770} {arXiv:2408.11770 [astro-ph.CO]}
  \BibitemShut {NoStop}%
\bibitem [{\citenamefont {Murakami}\ \emph {et~al.}(2023)\citenamefont
  {Murakami}, \citenamefont {Riess}, \citenamefont {Stahl}, \citenamefont
  {Kenworthy}, \citenamefont {Pluck}, \citenamefont {Macoretta}, \citenamefont
  {Brout}, \citenamefont {Jones}, \citenamefont {Scolnic},\ and\ \citenamefont
  {Filippenko}}]{Murakami:2023xuy}%
  \BibitemOpen
  \bibfield  {author} {\bibinfo {author} {\bibfnamefont {Y.~S.}\ \bibnamefont
  {Murakami}}, \bibinfo {author} {\bibfnamefont {A.~G.}\ \bibnamefont {Riess}},
  \bibinfo {author} {\bibfnamefont {B.~E.}\ \bibnamefont {Stahl}}, \bibinfo
  {author} {\bibfnamefont {W.~D.}\ \bibnamefont {Kenworthy}}, \bibinfo {author}
  {\bibfnamefont {D.-M.~A.}\ \bibnamefont {Pluck}}, \bibinfo {author}
  {\bibfnamefont {A.}~\bibnamefont {Macoretta}}, \bibinfo {author}
  {\bibfnamefont {D.}~\bibnamefont {Brout}}, \bibinfo {author} {\bibfnamefont
  {D.~O.}\ \bibnamefont {Jones}}, \bibinfo {author} {\bibfnamefont {D.~M.}\
  \bibnamefont {Scolnic}},\ and\ \bibinfo {author} {\bibfnamefont {A.~V.}\
  \bibnamefont {Filippenko}},\ }\bibfield  {title} {\bibinfo {title}
  {{Leveraging SN Ia spectroscopic similarity to improve the measurement of H
  $_{0}$}},\ }\href {https://doi.org/10.1088/1475-7516/2023/11/046} {\bibfield
  {journal} {\bibinfo  {journal} {JCAP}\ }\textbf {\bibinfo {volume} {11}},\
  \bibinfo {pages} {046}},\ \Eprint {https://arxiv.org/abs/2306.00070}
  {arXiv:2306.00070 [astro-ph.CO]} \BibitemShut {NoStop}%
\bibitem [{\citenamefont {Breuval}\ \emph {et~al.}(2024)\citenamefont
  {Breuval}, \citenamefont {Riess}, \citenamefont {Casertano}, \citenamefont
  {Yuan}, \citenamefont {Macri}, \citenamefont {Romaniello}, \citenamefont
  {Murakami}, \citenamefont {Scolnic}, \citenamefont {Anand},\ and\
  \citenamefont {Soszy\'nski}}]{Breuval:2024lsv}%
  \BibitemOpen
  \bibfield  {author} {\bibinfo {author} {\bibfnamefont {L.}~\bibnamefont
  {Breuval}}, \bibinfo {author} {\bibfnamefont {A.~G.}\ \bibnamefont {Riess}},
  \bibinfo {author} {\bibfnamefont {S.}~\bibnamefont {Casertano}}, \bibinfo
  {author} {\bibfnamefont {W.}~\bibnamefont {Yuan}}, \bibinfo {author}
  {\bibfnamefont {L.~M.}\ \bibnamefont {Macri}}, \bibinfo {author}
  {\bibfnamefont {M.}~\bibnamefont {Romaniello}}, \bibinfo {author}
  {\bibfnamefont {Y.~S.}\ \bibnamefont {Murakami}}, \bibinfo {author}
  {\bibfnamefont {D.}~\bibnamefont {Scolnic}}, \bibinfo {author} {\bibfnamefont
  {G.~S.}\ \bibnamefont {Anand}},\ and\ \bibinfo {author} {\bibfnamefont
  {I.}~\bibnamefont {Soszy\'nski}},\ }\bibfield  {title} {\bibinfo {title}
  {{Small Magellanic Cloud Cepheids Observed with the Hubble Space Telescope
  Provide a New Anchor for the SH0ES Distance Ladder}},\ }\href
  {https://doi.org/10.3847/1538-4357/ad630e} {\bibfield  {journal} {\bibinfo
  {journal} {Astrophys. J.}\ }\textbf {\bibinfo {volume} {973}},\ \bibinfo
  {pages} {30} (\bibinfo {year} {2024})},\ \Eprint
  {https://arxiv.org/abs/2404.08038} {arXiv:2404.08038 [astro-ph.CO]}
  \BibitemShut {NoStop}%
\bibitem [{\citenamefont {Abbott}\ \emph {et~al.}(2024)\citenamefont {Abbott}
  \emph {et~al.}}]{DES:2024jxu}%
  \BibitemOpen
  \bibfield  {author} {\bibinfo {author} {\bibfnamefont {T.~M.~C.}\
  \bibnamefont {Abbott}} \emph {et~al.} (\bibinfo {collaboration} {DES}),\
  }\bibfield  {title} {\bibinfo {title} {{The Dark Energy Survey: Cosmology
  Results with \ensuremath{\sim}1500 New High-redshift Type Ia Supernovae Using
  the Full 5 yr Data Set}},\ }\href {https://doi.org/10.3847/2041-8213/ad6f9f}
  {\bibfield  {journal} {\bibinfo  {journal} {Astrophys. J. Lett.}\ }\textbf
  {\bibinfo {volume} {973}},\ \bibinfo {pages} {L14} (\bibinfo {year}
  {2024})},\ \Eprint {https://arxiv.org/abs/2401.02929} {arXiv:2401.02929
  [astro-ph.CO]} \BibitemShut {NoStop}%
\bibitem [{\citenamefont {Rubin}\ \emph {et~al.}(2023)\citenamefont {Rubin}
  \emph {et~al.}}]{Rubin:2023ovl}%
  \BibitemOpen
  \bibfield  {author} {\bibinfo {author} {\bibfnamefont {D.}~\bibnamefont
  {Rubin}} \emph {et~al.},\ }\bibfield  {title} {\bibinfo {title} {{Union
  Through UNITY: Cosmology with 2,000 SNe Using a Unified Bayesian
  Framework}},\ }\href@noop {} {\  (\bibinfo {year} {2023})},\ \Eprint
  {https://arxiv.org/abs/2311.12098} {arXiv:2311.12098 [astro-ph.CO]}
  \BibitemShut {NoStop}%
\bibitem [{\citenamefont {de~Putter}\ and\ \citenamefont
  {Linder}(2008)}]{dePutter:2008wt}%
  \BibitemOpen
  \bibfield  {author} {\bibinfo {author} {\bibfnamefont {R.}~\bibnamefont
  {de~Putter}}\ and\ \bibinfo {author} {\bibfnamefont {E.~V.}\ \bibnamefont
  {Linder}},\ }\bibfield  {title} {\bibinfo {title} {{Calibrating Dark
  Energy}},\ }\href {https://doi.org/10.1088/1475-7516/2008/10/042} {\bibfield
  {journal} {\bibinfo  {journal} {JCAP}\ }\textbf {\bibinfo {volume} {10}},\
  \bibinfo {pages} {042}},\ \Eprint {https://arxiv.org/abs/0808.0189}
  {arXiv:0808.0189 [astro-ph]} \BibitemShut {NoStop}%
\bibitem [{\citenamefont {Wolf}\ and\ \citenamefont
  {Ferreira}(2023)}]{Wolf:2023uno}%
  \BibitemOpen
  \bibfield  {author} {\bibinfo {author} {\bibfnamefont {W.~J.}\ \bibnamefont
  {Wolf}}\ and\ \bibinfo {author} {\bibfnamefont {P.~G.}\ \bibnamefont
  {Ferreira}},\ }\bibfield  {title} {\bibinfo {title} {{Underdetermination of
  dark energy}},\ }\href {https://doi.org/10.1103/PhysRevD.108.103519}
  {\bibfield  {journal} {\bibinfo  {journal} {Phys. Rev. D}\ }\textbf {\bibinfo
  {volume} {108}},\ \bibinfo {pages} {103519} (\bibinfo {year} {2023})},\
  \Eprint {https://arxiv.org/abs/2310.07482} {arXiv:2310.07482 [astro-ph.CO]}
  \BibitemShut {NoStop}%
\bibitem [{\citenamefont {Aghanim}\ \emph
  {et~al.}(2020{\natexlab{b}})\citenamefont {Aghanim} \emph {et~al.}}]{PL2018}%
  \BibitemOpen
  \bibfield  {author} {\bibinfo {author} {\bibfnamefont {N.}~\bibnamefont
  {Aghanim}} \emph {et~al.} (\bibinfo {collaboration} {Planck}),\ }\bibfield
  {title} {\bibinfo {title} {{Planck 2018 results. VIII. Gravitational
  lensing}},\ }\href {https://doi.org/10.1051/0004-6361/201833886} {\bibfield
  {journal} {\bibinfo  {journal} {\aap}\ }\textbf {\bibinfo {volume} {641}},\
  \bibinfo {pages} {A8} (\bibinfo {year} {2020}{\natexlab{b}})},\ \Eprint
  {https://arxiv.org/abs/1807.06210} {arXiv:1807.06210 [astro-ph.CO]}
  \BibitemShut {NoStop}%
\bibitem [{\citenamefont {Adame}\ \emph {et~al.}(2024)\citenamefont {Adame}
  \emph {et~al.}}]{DESI:2024hhd}%
  \BibitemOpen
  \bibfield  {author} {\bibinfo {author} {\bibfnamefont {A.~G.}\ \bibnamefont
  {Adame}} \emph {et~al.} (\bibinfo {collaboration} {DESI}),\ }\bibfield
  {title} {\bibinfo {title} {{DESI 2024 VII: Cosmological Constraints from the
  Full-Shape Modeling of Clustering Measurements}},\ }\href@noop {} {\
  (\bibinfo {year} {2024})},\ \Eprint {https://arxiv.org/abs/2411.12022}
  {arXiv:2411.12022 [astro-ph.CO]} \BibitemShut {NoStop}%
\bibitem [{\citenamefont {Akrami}\ \emph {et~al.}(2020)\citenamefont {Akrami}
  \emph {et~al.}}]{Akrami:2020bpw}%
  \BibitemOpen
  \bibfield  {author} {\bibinfo {author} {\bibfnamefont {Y.}~\bibnamefont
  {Akrami}} \emph {et~al.} (\bibinfo {collaboration} {Planck}),\ }\bibfield
  {title} {\bibinfo {title} {{$Planck$ intermediate results. LVII. Joint Planck
  LFI and HFI data processing}},\ }\href
  {https://doi.org/10.1051/0004-6361/202038073} {\bibfield  {journal} {\bibinfo
   {journal} {\aap}\ }\textbf {\bibinfo {volume} {643}},\ \bibinfo {pages}
  {A42} (\bibinfo {year} {2020})},\ \Eprint {https://arxiv.org/abs/2007.04997}
  {arXiv:2007.04997 [astro-ph.CO]} \BibitemShut {NoStop}%
\bibitem [{\citenamefont {Rosenberg}\ \emph {et~al.}(2022)\citenamefont
  {Rosenberg}, \citenamefont {Gratton},\ and\ \citenamefont
  {Efstathiou}}]{Rosenberg:2022sdy}%
  \BibitemOpen
  \bibfield  {author} {\bibinfo {author} {\bibfnamefont {E.}~\bibnamefont
  {Rosenberg}}, \bibinfo {author} {\bibfnamefont {S.}~\bibnamefont {Gratton}},\
  and\ \bibinfo {author} {\bibfnamefont {G.}~\bibnamefont {Efstathiou}},\
  }\bibfield  {title} {\bibinfo {title} {{CMB power spectra and cosmological
  parameters from Planck PR4 with CamSpec}},\ }\href
  {https://doi.org/10.1093/mnras/stac2744} {\bibfield  {journal} {\bibinfo
  {journal} {Mon. Not. Roy. Astron. Soc.}\ }\textbf {\bibinfo {volume} {517}},\
  \bibinfo {pages} {4620} (\bibinfo {year} {2022})},\ \Eprint
  {https://arxiv.org/abs/2205.10869} {arXiv:2205.10869 [astro-ph.CO]}
  \BibitemShut {NoStop}%
\bibitem [{\citenamefont {Torrado}\ and\ \citenamefont
  {Lewis}(2021)}]{Torrado:2020dgo}%
  \BibitemOpen
  \bibfield  {author} {\bibinfo {author} {\bibfnamefont {J.}~\bibnamefont
  {Torrado}}\ and\ \bibinfo {author} {\bibfnamefont {A.}~\bibnamefont
  {Lewis}},\ }\bibfield  {title} {\bibinfo {title} {{Cobaya: Code for Bayesian
  Analysis of hierarchical physical models}},\ }\href
  {https://doi.org/10.1088/1475-7516/2021/05/057} {\bibfield  {journal}
  {\bibinfo  {journal} {\jcap}\ }\textbf {\bibinfo {volume} {05}},\ \bibinfo
  {pages} {057} (\bibinfo {year} {2021})},\ \Eprint
  {https://arxiv.org/abs/2005.05290} {arXiv:2005.05290 [astro-ph.IM]}
  \BibitemShut {NoStop}%
\bibitem [{\citenamefont {Lewis}\ \emph {et~al.}(2000)\citenamefont {Lewis},
  \citenamefont {Challinor},\ and\ \citenamefont {Lasenby}}]{Lewis:1999bs}%
  \BibitemOpen
  \bibfield  {author} {\bibinfo {author} {\bibfnamefont {A.}~\bibnamefont
  {Lewis}}, \bibinfo {author} {\bibfnamefont {A.}~\bibnamefont {Challinor}},\
  and\ \bibinfo {author} {\bibfnamefont {A.}~\bibnamefont {Lasenby}},\
  }\bibfield  {title} {\bibinfo {title} {{Efficient computation of CMB
  anisotropies in closed FRW models}},\ }\href {https://doi.org/10.1086/309179}
  {\bibfield  {journal} {\bibinfo  {journal} {\apj}\ }\textbf {\bibinfo
  {volume} {538}},\ \bibinfo {pages} {473} (\bibinfo {year} {2000})},\ \Eprint
  {https://arxiv.org/abs/astro-ph/9911177} {arXiv:astro-ph/9911177 [astro-ph]}
  \BibitemShut {NoStop}%
\bibitem [{\citenamefont {Lewis}(2019)}]{Lewis:2019xzd}%
  \BibitemOpen
  \bibfield  {author} {\bibinfo {author} {\bibfnamefont {A.}~\bibnamefont
  {Lewis}},\ }\bibfield  {title} {\bibinfo {title} {{GetDist: a Python package
  for analysing Monte Carlo samples}},\ }\href@noop {} {\  (\bibinfo {year}
  {2019})},\ \bibinfo {note} {{\url{https://getdist.readthedocs.io}}},\ \Eprint
  {https://arxiv.org/abs/1910.13970} {arXiv:1910.13970 [astro-ph.IM]}
  \BibitemShut {NoStop}%
\bibitem [{\citenamefont {Gialamas}\ \emph {et~al.}(2025)\citenamefont
  {Gialamas}, \citenamefont {H\"utsi}, \citenamefont {Kannike}, \citenamefont
  {Racioppi}, \citenamefont {Raidal}, \citenamefont {Vasar},\ and\
  \citenamefont {Veerm\"ae}}]{Gialamas:2024lyw}%
  \BibitemOpen
  \bibfield  {author} {\bibinfo {author} {\bibfnamefont {I.~D.}\ \bibnamefont
  {Gialamas}}, \bibinfo {author} {\bibfnamefont {G.}~\bibnamefont {H\"utsi}},
  \bibinfo {author} {\bibfnamefont {K.}~\bibnamefont {Kannike}}, \bibinfo
  {author} {\bibfnamefont {A.}~\bibnamefont {Racioppi}}, \bibinfo {author}
  {\bibfnamefont {M.}~\bibnamefont {Raidal}}, \bibinfo {author} {\bibfnamefont
  {M.}~\bibnamefont {Vasar}},\ and\ \bibinfo {author} {\bibfnamefont
  {H.}~\bibnamefont {Veerm\"ae}},\ }\bibfield  {title} {\bibinfo {title}
  {{Interpreting DESI 2024 BAO: Late-time dynamical dark energy or a local
  effect?}},\ }\href {https://doi.org/10.1103/PhysRevD.111.043540} {\bibfield
  {journal} {\bibinfo  {journal} {Phys. Rev. D}\ }\textbf {\bibinfo {volume}
  {111}},\ \bibinfo {pages} {043540} (\bibinfo {year} {2025})},\ \Eprint
  {https://arxiv.org/abs/2406.07533} {arXiv:2406.07533 [astro-ph.CO]}
  \BibitemShut {NoStop}%
\bibitem [{\citenamefont {{Efstathiou}}(2025)}]{Efstathiou:2024xcq}%
  \BibitemOpen
  \bibfield  {author} {\bibinfo {author} {\bibfnamefont {G.}~\bibnamefont
  {{Efstathiou}}},\ }\bibfield  {title} {\bibinfo {title} {{Evolving dark
  energy or supernovae systematics?}},\ }\href
  {https://doi.org/10.1093/mnras/staf301} {\bibfield  {journal} {\bibinfo
  {journal} {\mnras}\ }\textbf {\bibinfo {volume} {538}},\ \bibinfo {pages}
  {875} (\bibinfo {year} {2025})},\ \Eprint {https://arxiv.org/abs/2408.07175}
  {arXiv:2408.07175 [astro-ph.CO]} \BibitemShut {NoStop}%
\bibitem [{\citenamefont {Dhawan}\ \emph {et~al.}(2024)\citenamefont {Dhawan},
  \citenamefont {Popovic},\ and\ \citenamefont {Goobar}}]{Dhawan:2024gqy}%
  \BibitemOpen
  \bibfield  {author} {\bibinfo {author} {\bibfnamefont {S.}~\bibnamefont
  {Dhawan}}, \bibinfo {author} {\bibfnamefont {B.}~\bibnamefont {Popovic}},\
  and\ \bibinfo {author} {\bibfnamefont {A.}~\bibnamefont {Goobar}},\
  }\bibfield  {title} {\bibinfo {title} {{The axis of systematic bias in SN Ia
  cosmology and implications for DESI 2024 results}},\ }\href@noop {} {\
  (\bibinfo {year} {2024})},\ \Eprint {https://arxiv.org/abs/2409.18668}
  {arXiv:2409.18668 [astro-ph.CO]} \BibitemShut {NoStop}%
\bibitem [{\citenamefont {Notari}\ \emph {et~al.}(2024)\citenamefont {Notari},
  \citenamefont {Redi},\ and\ \citenamefont {Tesi}}]{Notari:2024zmi}%
  \BibitemOpen
  \bibfield  {author} {\bibinfo {author} {\bibfnamefont {A.}~\bibnamefont
  {Notari}}, \bibinfo {author} {\bibfnamefont {M.}~\bibnamefont {Redi}},\ and\
  \bibinfo {author} {\bibfnamefont {A.}~\bibnamefont {Tesi}},\ }\bibfield
  {title} {\bibinfo {title} {{BAO vs. SN evidence for evolving dark energy}},\
  }\href@noop {} {\  (\bibinfo {year} {2024})},\ \Eprint
  {https://arxiv.org/abs/2411.11685} {arXiv:2411.11685 [astro-ph.CO]}
  \BibitemShut {NoStop}%
\bibitem [{\citenamefont {Vincenzi}\ \emph {et~al.}(2025)\citenamefont
  {Vincenzi} \emph {et~al.}}]{DES:2025tir}%
  \BibitemOpen
  \bibfield  {author} {\bibinfo {author} {\bibfnamefont {M.}~\bibnamefont
  {Vincenzi}} \emph {et~al.} (\bibinfo {collaboration} {DES}),\ }\bibfield
  {title} {\bibinfo {title} {{Comparing the DES-SN5YR and Pantheon+ SN
  cosmology analyses: Investigation based on ''Evolving Dark Energy or
  Supernovae systematics?''}},\ }\href@noop {} {\  (\bibinfo {year} {2025})},\
  \Eprint {https://arxiv.org/abs/2501.06664} {arXiv:2501.06664 [astro-ph.CO]}
  \BibitemShut {NoStop}%
\bibitem [{\citenamefont {Huang}\ \emph {et~al.}(2025)\citenamefont {Huang},
  \citenamefont {Cai},\ and\ \citenamefont {Wang}}]{Huang:2025som}%
  \BibitemOpen
  \bibfield  {author} {\bibinfo {author} {\bibfnamefont {L.}~\bibnamefont
  {Huang}}, \bibinfo {author} {\bibfnamefont {R.-G.}\ \bibnamefont {Cai}},\
  and\ \bibinfo {author} {\bibfnamefont {S.-J.}\ \bibnamefont {Wang}},\
  }\bibfield  {title} {\bibinfo {title} {{The DESI 2024 hint for dynamical dark
  energy is biased by low-redshift supernovae}},\ }\href@noop {} {\  (\bibinfo
  {year} {2025})},\ \Eprint {https://arxiv.org/abs/2502.04212}
  {arXiv:2502.04212 [astro-ph.CO]} \BibitemShut {NoStop}%
\bibitem [{\citenamefont {Brout}\ \emph {et~al.}(2022)\citenamefont {Brout}
  \emph {et~al.}}]{Brout:2022vxf}%
  \BibitemOpen
  \bibfield  {author} {\bibinfo {author} {\bibfnamefont {D.}~\bibnamefont
  {Brout}} \emph {et~al.},\ }\bibfield  {title} {\bibinfo {title} {{The
  Pantheon+ Analysis: Cosmological Constraints}},\ }\href
  {https://doi.org/10.3847/1538-4357/ac8e04} {\bibfield  {journal} {\bibinfo
  {journal} {Astrophys. J.}\ }\textbf {\bibinfo {volume} {938}},\ \bibinfo
  {pages} {110} (\bibinfo {year} {2022})},\ \Eprint
  {https://arxiv.org/abs/2202.04077} {arXiv:2202.04077 [astro-ph.CO]}
  \BibitemShut {NoStop}%
\bibitem [{\citenamefont {Kaplinghat}\ and\ \citenamefont
  {Bridle}(2005)}]{Kaplinghat:2003vf}%
  \BibitemOpen
  \bibfield  {author} {\bibinfo {author} {\bibfnamefont {M.}~\bibnamefont
  {Kaplinghat}}\ and\ \bibinfo {author} {\bibfnamefont {S.}~\bibnamefont
  {Bridle}},\ }\bibfield  {title} {\bibinfo {title} {{Testing for a
  superacceleration phase of the universe}},\ }\href
  {https://doi.org/10.1103/PhysRevD.71.123003} {\bibfield  {journal} {\bibinfo
  {journal} {Phys. Rev. D}\ }\textbf {\bibinfo {volume} {71}},\ \bibinfo
  {pages} {123003} (\bibinfo {year} {2005})},\ \Eprint
  {https://arxiv.org/abs/astro-ph/0312430} {arXiv:astro-ph/0312430}
  \BibitemShut {NoStop}%
\bibitem [{\citenamefont {Shajib}\ and\ \citenamefont
  {Frieman}(2025)}]{Shajib:2025tpd}%
  \BibitemOpen
  \bibfield  {author} {\bibinfo {author} {\bibfnamefont {A.~J.}\ \bibnamefont
  {Shajib}}\ and\ \bibinfo {author} {\bibfnamefont {J.~A.}\ \bibnamefont
  {Frieman}},\ }\href@noop {} {\bibinfo {title} {{Evolving dark energy models:
  Current and forecast constraints}}} (\bibinfo {year} {2025}),\ \Eprint
  {https://arxiv.org/abs/2502.06929} {arXiv:2502.06929 [astro-ph.CO]}
  \BibitemShut {NoStop}%
\end{thebibliography}
\end{document}